\documentclass[graybox]{svmult}

\usepackage{type1cm}        
\usepackage{makeidx}         
\usepackage{graphicx}        
\usepackage{multicol}        
\usepackage[bottom]{footmisc}

\usepackage{newtxtext}       %
\usepackage[varvw]{newtxmath}       

\makeindex             

\usepackage{url}

\usepackage{enumitem}

\newcommand{\map}{\mathfrak{m}}
\newcommand{\tree}{\mathfrak{t}}
\newcommand{\vertices}{\mathsf{V}}
\newcommand{\edges}{\mathsf{E}}
\newcommand{\faces}{\mathsf{F}}

\newcommand{\ind}{\mathbf{1}}
\newcommand{\rmd}{\mathrm{d}}

\begin{document}

\title*{Lessons from the Mathematics of Two-Dimensional Euclidean Quantum Gravity}
\titlerunning{Two-Dimensional Euclidean Quantum Gravity}
\author{Timothy Budd}
\institute{Timothy Budd \at Radboud University, Nijmegen, The Netherlands, \email{t.budd@science.ru.nl}}
%
%
\maketitle

\abstract{
The search for a mathematical foundation for the path integral of Euclidean quantum gravity calls for the construction of random geometry on the spacetime manifold.
Following developments in physics on the two-dimensional theory, random geometry on the 2-sphere has in recent years received much attention in the mathematical literature, which has led to a fully rigorous implementation of the path integral formulation of two-dimensional Euclidean quantum gravity.
In this chapter we review several important mathematical developments that may serve as guiding principles for approaching Euclidean quantum gravity in dimensions higher than two.
Our starting point is the discrete geometry encoded by random planar maps, which realizes a lattice discretization of the path integral.
We recap the enumeration of planar maps via their generating functions and show how bijections with trees explain the surprising simplicity of some of these.
Then we explain how to handle infinite planar maps and to analyze their exploration via the peeling process.
The aforementioned trees provide the basis for the construction of the universal continuum limit of the random discrete geometries, known as the Brownian sphere, which represents the random geometry underlying two-dimensional Euclidean quantum gravity in the absence of matter.
}

\keywords{2D Quantum Gravity, Random Geometry, Random Planar Maps, Tree Bijections, Scaling Limits, Brownian Sphere, Peeling Process}

\section{Introduction}

\subsection{Euclidean Quantum Gravity}

The path integral approach to Euclidean Quantum Gravity \cite{Gibbons_Action_1977,Gibbons_Euclidean_1993} aims at assigning a mathematical meaning to a functional integral of the form
\begin{equation}\label{eq:pathintegral}
    \mathcal{Z} = \int \mathcal{D}[g_{ab}] e^{-S[g_{ab}]},
\end{equation}
where the integration is over all isometry classes $[g_{ab}]$ of Riemannian metrics $g_{ab}$ on a $d$-dimensional spacetime manifold $M$ with an appropriate action $S[g_{ab}]$, like the (Euclidean) Einstein--Hilbert action 
\begin{equation}\label{eq:einsteinhilbert}
    S[g_{ab}] = \frac{1}{16\pi G_{\mathrm{N}}} \int_M \rmd^d x\sqrt{g} (-R + 2\Lambda).
\end{equation}
It has the key advantage over the Feynman path integral of Lorentzian quantum gravity that, at least formally, the integrand is real and positive, allowing for an interpretation of $\mathcal{Z}$ as a partition function of a statistical system in which $e^{-S[g_{ab}]}$ takes on the role of the Boltzmann weight assigned to the geometry $[g_{ab}]$.
Putting aside the matter of relating the two path integrals via a ``Wick rotation'', it turns the problem of Euclidean quantum gravity into a question of probability theory: does there exist a suitable probability measure 
\begin{equation}\label{eq:qgmeasure}
    \frac{1}{\mathcal{Z}}\mathcal{D}[g_{ab}] e^{-S[g_{ab}]}
\end{equation}
on the space of geometries on $M$?

Needless to say, there are significant challenges on the way to answering this question.
As with other (Euclidean) Quantum Field Theory any approximation scheme of the integral \eqref{eq:pathintegral} will encounter both ultraviolet and infrared divergencies that have to be suitably regularized.
The absence of a background geometry, peculiar to gravity, makes categorizing the field degrees of freedom by scale (and thus deciding which are infrared or ultraviolet) already a non-trivial affair, because a notion of scale necessarily involves the dynamical metric itself.
In addition, since gravity is perturbatively non-renormalizable, one should be prepared to deal with metric degrees of freedom at short length scales (i.e.\ in the ultraviolet) that are strongly interacting.
Based on this one may question whether the space of Riemannian metrics provides an arena general enough to host the sought-after probability measure, since many classical notions of Riemannian geometry (like curvature and geodesics) may be lost when the metric becomes too irregular, for instance if the metric tensor is non-differentiable or even distributional.
It may thus be necessary to find an appropriate generalization of the Riemannian metric that can support such a measure.
Finally, in constructing the measure one should overcome the issue that the Einstein--Hilbert action (for $d\geq 3$) is unbounded from below, suggesting that certain highly curved geometries will receive much higher Boltzmann weight than the classical Einstein solutions (known as \emph{conformal factor problem} \cite{Gibbons_Path_1978,Mazur_path_1990}).

A mathematically precise construction of the partition function has not been achieved yet for any manifold of dimension $d\geq 3$, but significant progress is reported elsewhere in the handbook. 
Notably, renormalization group methods \cite{Reuter_Nonperturbative_1998,Dou_running_1998,Reuter_Renormalization_2002} (see \cite{Reuter_Quantum_2019} for a thorough account) have found indications that Euclidean Quantum Gravity is \emph{asymptotically safe} \cite{Weinberg_Ultraviolet_1979}, meaning that it is renormalizable with an interacting fixed point in the ultraviolet.
Lattice discretization approaches like Euclidean Dynamical Triangulations \cite{Boulatov_phase_1991,Ambjoern_Four_1992,Ambjoern_vacuum_1992,Agishtein_Three_1991} and Causal Dynamical Triangulations \cite{Ambjoern_Nonperturbative_2012,Loll_Quantum_2019} provide a complementary perspective based on numerical methods.

In this chapter, however, we focus on Euclidean Quantum Gravity on the two-dimensional sphere which admits a fully rigorous probabilistic interpretation.
It has been long known in the physics literature that the partition function in two dimensions is susceptible to analytic computation from various starting points, including the lattice approaches via Dynamical Triangulations \cite{David_Planar_1985,Ambjorn_Diseases_1985,Kazakov_Critical_1985,Ambjoern_Multiloop_1990,Ambjorn_Quantum_1997} and matrix models \cite{Hooft_planar_1993,Brezin_Planar_1978,DiFrancesco_2D_1995} as well as conformal field theory approaches via Liouville field theory 
\cite{Polyakov1981,Knizhnik_Fractal_1988,David_Conformal_1988,Distler_Conformal_1989}.
The focus of this chapter, however, is on the developments in the mathematical literature in the last two decades that have put these computations on a rigorous footing and have culminated in an unambiguous construction of the probability measure \eqref{eq:qgmeasure} representing two-dimensional Euclidean quantum gravity.

Of course, it is a greatly simplified toy model compared to quantum gravity on more realistic four-dimensional manifolds, but one that is far from trivial and already requires us to depart from certain classical intuition coming from Riemannian geometry.
It thus forms an important test bed for our mathematical methods and several lessons can be learned (at least on what \emph{not} to take for granted when searching for higher-dimensional analogues).

\subsection{Two-dimensional quantum gravity}

One aspect which sets gravity in two dimensions apart from its higher-dimensional counterparts is that Einstein's field equations in vacuum are trivial: every Riemannian metric is a solution when $\Lambda=0$ and none is when $\Lambda\neq 0$. 
This is tied to the fact that the curvature integral in the Einstein--Hilbert action \eqref{eq:einsteinhilbert} for $d=2$ is a topological invariant due to the Gauss--Bonnet formula, so fixing the manifold $M = S^2$ to be the 2-sphere the only dependence on the metric is through its total volume $\int_{S^2} \sqrt{g}$.
The partition function \eqref{eq:pathintegral} can therefore formally be recast as an ordinary integration over the volume $V$ of the \emph{canonical partition function} $\mathcal{Z}_V$,
\begin{equation}\label{eq:partitionfunction}
    \mathcal{Z}_V = \int \mathcal{D}[g_{ab}] \delta\left(V - \int_{S^2} \sqrt{g}\right), \qquad \mathcal{Z} = \int_0^\infty \rmd V\,e^{-\frac{\Lambda}{16\pi G_{\mathrm{N}}} V} \mathcal{Z}_V.
\end{equation} 
Since every geometry of volume $V$ receives the same Boltzmann weight, the probability measure of two-dimensional quantum gravity (at fixed volume) should amount to a suitable notion of sampling a metric on $S^2$ \emph{uniformly} at random.

It is not at all obvious how to interpret this in the infinite-dimensional space of Riemannian geometries on $S^2$, but two-dimensional \emph{Euclidean Dynamical Triangulations} (EDT) provides a natural lattice discretization \cite{David_Planar_1985,Ambjorn_Diseases_1985,Kazakov_Critical_1985,Ambjoern_Multiloop_1990,Ambjorn_Quantum_1997}.
Instead of considering the full set of Riemannian geometries on $S^2$, one restricts to the piece-wise flat geometries that can be assembled from a fixed number of equilateral Euclidean triangles of identical size. 
This introduces both an ultraviolet cutoff, by having a finite lattice spacing, and an infrared cutoff, by limiting the maximal diameter of the geometry.
Since the set of geometries is now finite, one can easily select a uniform random metric by assigning equal probability to each.
Then the hope is that this probability measure admits a well-defined continuum limit upon shrinking the triangles while increasing their number.
We will review this limit, known as the Brownian sphere, in detail in the mathematical framework of random planar maps, where informally the building blocks are arbitrary regular (but mostly even-sided) polygons with unit side length.

Based on the extensive mathematical literature we can summarize some important lessons as follows:
\begin{itemize}
    \item The universality observed in enumeration formulas for planar maps can be understood combinatorially via the existence of bijections between maps and trees (Section~\ref{sec:treebijections}).
    \item The infrared cutoff in the probability measure can be consistently removed by considering the limit of random infinite planar maps in an appropriate topology, known as the local topology (Section~\ref{sec:locallimits}). Often the random infinite geometry is easier to analyze than one of fixed finite size, for instance when studying explorations (Section~\ref{sec:peeling}).
    \item The ultraviolet cutoff can be removed via a continuum limit in which the lattice spacing scales appropriately with the size of the random planar map (Section~\ref{sec:continuumlimit}). 
    The convergence takes place with respect to the Gromov--Hausdorff topology on the space of (compact) metric spaces (sets equipped with distance functions), which is significantly larger than the space of Riemannian geometries.
    The limit, known as the Brownian sphere, is a random metric space with the topology of $S^2$ and well-defined notion of geodesics, providing a precise realization of the probability measure in \eqref{eq:qgmeasure}.
    However, it is not Riemannian as becomes apparent when examining its geodesics more closely (Section~\ref{sec:Browniansphereproperties}).
    \item Removing both cutoffs naturally leads to a random metric space, known as the Brownian plane (Section~\ref{sec:locallimits}), with exact scaling symmetry, in the sense that multiplying all distances by a positive constant does not change its distribution. Such a scale-invariant random geometry should be interpreted as realizing a fixed point of the renormalization group associated to Euclidean quantum gravity.
\end{itemize}

\section{Planar maps and their enumeration}

Before delving into random geometries and their properties, we will discuss in this section how discrete surfaces are conveniently encoded in terms of maps and how one can approach their enumeration.

\subsection{Maps as discrete surfaces}

In the previous section we informally introduced discrete surfaces as  two-dimensional Riemannian geometries that can be obtained from gluing together regular Euclidean polygons. 
But since we are interested in precise enumeration, it is important to choose the combinatorial representation in an unambiguous fashion.

\begin{figure}[t]
    \sidecaption[t]
    \includegraphics[width=.6\linewidth]{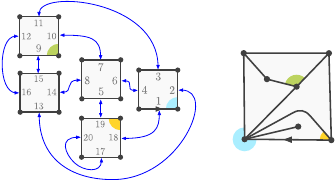}
    \caption{A planar map as a gluing of polygons and as an embedded graph. This example shows a (rooted) quadrangulation with $5$ faces, $10$ edges and $7$ vertices. Colors have been added to guide the eye.\label{fig:planarmap}}
\end{figure}

A simple way to do so is to start with a finite set of regular polygons of unit side length and label the sides by integers $1, 2, \ldots, 2n$ in an arbitrary fashion. 
Given a \emph{matching} of $\{1,\ldots,2n\}$, i.e.\ a partition of $\{1,\ldots,2n\}$ into pairs, one may construct a closed surface by gluing the $2n$ sides accordingly, see Fig.~\ref{fig:planarmap}.
More precisely, we assume that the polygons have an orientation and that their sides, also referred to as \emph{half-edges}, are oriented in counterclockwise direction.
We can then make the gluing operation unambiguous by requiring that pairs of half-edges are identified with opposite orientation.
If the resulting geometry is connected, this gluing of polygons is called a \emph{map}.

Observe that the $2n$ half-edges of a map are identified into a graph with $n$ edges (hence \emph{half}-edges) that is embedded in a topological surface, which is necessarily orientable and determined by its genus.
This brings us to an alternative definition of a map as a graph together with a proper embedding in a closed oriented surface, where by \emph{proper embedding} we mean that the edges do not intersect themselves or other edges, except where they meet at vertices, and that the edges together delimit a collection of topological disks, called the \emph{faces} of the map (which are nothing but the interiors of the polygons above). 
One then views two maps as equivalent if they can be related via an orientation-preserving homeomorphism of the surface.
We should remark that graphs (and thus maps) are allowed to have more than one edge between a pair of vertices and to have edges starting and ending at the same vertex.

The sets of vertices, edges and faces of a map $\map$ are denoted $\vertices(\map)$, $\edges(\map)$ and $\faces(\map)$ respectively.
According to Euler's formula, the numbers of vertices, edges and faces are related by
\begin{equation}\label{eq:eulerformula}
    |\vertices(\map)| - |\edges(\map)| + |\faces(\map)| = 2-2g,
\end{equation}
where $g$ is the genus of the corresponding surface.
In the case $g=0$ we are dealing with a \emph{planar map}, i.e.\ a polygonal gluing into a topological sphere or, equivalently, a planar graph properly embedded in the sphere.
We take the \emph{degree} of a vertex or face to be the number of edges incident to it (where we count an edge twice if both endpoints are at the same vertex or both sides adjacent to the same face). 
Special families of maps that we will encounter are \emph{triangulations}\footnote{In the literature these triangulations are sometimes called \emph{type I}, because loops and multiple edges between vertices are allowed. \emph{Loopless} triangulations, which forbid loops but allow multiple edges, are \emph{type II}. \emph{Simple} triangulations forbid both and are \emph{type III}.}, \emph{quadrangulations}, and \emph{even maps}, whose faces all have degree three, degree four or even degree respectively.
Even planar maps are also known as \emph{bipartite planar maps}, because the vertices of such a map can be colored with two colors such that the endpoints of each edge have different color.
This should be contrasted with the case of even maps of genus $g\geq 1$, which are not necessarily bipartite.

It should be observed that the polygonal gluing provides a highly redundant description of a map, since the half-edges carry an arbitrary labeling, whereas a description in terms of unlabeled embedded graphs is often too abstract to work with. 
This is especially true when the map in question possesses internal symmetries, i.e.\ when there exist orientation-preserving homeomorphisms of the sphere that non-trivially permute the edges and vertices of the map, called \emph{automorphisms}.
In such a case the number of labeled maps corresponding to the same unlabeled map $\map$ is dependent on the size of the automorphism group $\operatorname{Aut}(\map)$ and is given by $(2n)! / |\operatorname{Aut}(\map)|$.
A practical middle ground is to consider unlabeled but \emph{rooted} maps, meaning that each map comes with a distinguished oriented edge (i.e.\ a distinguished half-edge).
Sending a labeled map to the (unlabeled) rooted map obtained by distinguishing its half-edge with label $1$ is precisely $(2n-1)!$-to-$1$, so the enumeration of both types are more easily related to each other.

\subsection{Random planar map models}

We are now ready to formulate in a precise combinatorial way the partition function of \emph{two-dimensional Euclidean Dynamical Triangulations} (EDT) as a lattice discretization of the partition function \eqref{eq:partitionfunction}.
It can be defined as a summation over all unlabeled, labeled or rooted planar triangulations as 
\begin{align} 
    Z_{\text{EDT}}(q_3) &= \sum_{\substack{\text{unlabeled planar}\\\text{triangulations }\map}} \frac{q_3^{|\faces(\map)|}}{|\operatorname{Aut}(\map)|} = \sum_{\substack{\text{labeled planar}\\\text{triangulations }\map}} \frac{q_3^{|\faces(\map)|}}{(2|\edges(\map)|)!} \\
    &= \sum_{\substack{\text{rooted planar}\\\text{triangulations }\map}} \frac{q_3^{|\faces(\map)|}}{2|\edges(\map)|} = \sum_{k=1}^\infty \frac{q_3^{2k}}{6k} T_k,
\end{align}
where $q_3$ can be interpreted as the exponential of the lattice cosmological constant and $T_k$ is the number of rooted planar triangulations with $2k$ triangles.
In particular $3q_3 Z_{\text{EDT}}'(q_3) = \sum_{k=1}^\infty q_3^{2k}T_k$ is nothing but the generating function of these numbers $T_k$, so its stands to reason that the enumeration of maps is at the heart of the model.
If $q_3>0$ is small enough that the sum converges, the Boltzmann weights define a probability distribution on triangulations known as the \emph{Boltzmann triangulation}. 
Furthermore, $T_k$ essentially is the corresponding canonical partition function of triangulations of fixed size $2k$, which therefore describes a random triangulation known as the \emph{uniform (rooted) triangulation} of size $2k$, meaning that each rooted triangulation with $2k$ triangles occurs with equal probability $1 / T_k$.

One may generalize this model to maps with faces of arbitrary degree by introducing a sequence of (non-negative) weights $\mathbf{q} = (q_1,q_2,q_3,\ldots)$ and assigning those to the faces according to their degree as well as a weight $t>0$ to each vertex.
Denoting the space of all rooted planar maps by $\mathcal{M}$, we are thus considering the partition function\footnote{Since the sum is over rooted planar maps, this partition function generalizes $3 q_3 Z_{\text{EDT}}'(q_3)$ rather than the partition function $Z_{\text{EDT}}(q_3)$ of unlabeled triangulations.}
\begin{align}
    Z(t,\mathbf{q}) = \sum_{\map \in \mathcal{M}} t^{|\vertices(\map)|} \prod_{f \in \faces(\map)} q_{\deg f}. \label{eq:boltzpartitionfunction}
\end{align}
If $Z(t,\mathbf{q}) < \infty$, we can normalize the summand by $1/Z(t,\mathbf{q})$ and take it to define a probability distribution on $\mathcal{M}$, which is called the \emph{$(t,\mathbf{q})$-Boltzmann planar map}.
The parameter $t$ actually is redundant here, because by Euler's formula \eqref{eq:eulerformula} we have
\begin{equation}
    Z(t,\mathbf{q}) = t^2\sum_{\map\in\mathcal{M}} \prod_{f\in\faces(\map)} t^{-1+\tfrac12\deg f}q_{\deg f} = t^2 Z(1,\tilde{\mathbf{q}}),\quad \tilde{q}_k \coloneqq t^{-1+\frac12k}q_k,
\end{equation}
meaning that the $(t,\mathbf{q})$-Boltzmann planar map is the same as the $\tilde{\mathbf{q}}$-Boltzmann planar map (with $t=1$).
For combinatorial reasons it can be useful to keep the parameter $t$, while we will often set $t=1$ later without loss of generality.

We could have chosen to assign weights to the vertices depending on their degrees instead of the faces, but the resulting models are related by duality. 
Here the \emph{dual} of a genus-$g$ map $\map$ is the map $\map^\dagger$ obtained by interchanging the roles of vertices and faces of $\map$, while keeping the same incidence relations.
More operationally, one places a vertex of $\map^\dagger$ in each face of $\map$ and one connects these by drawing an edge of $\map^\dagger$ intersecting each edge of $\map$.
The root of $\map^\dagger$ is taken to be the oriented edge starting at the root face and crossing the root of $\map$.
Since this is a bijection from $\mathcal{M}$ to itself and the vertex degrees of $\map^\dagger$ agree with the face degrees of $\map$, the dual of a $\mathbf{q}$-Boltzmann planar map is distributed according to the model with vertex weights.
A hybrid version, in which both vertices and faces receive weights, poses significant additional challenges, and only limited progress has been made towards solving such models (see \cite{DiFrancesco1993,Kazakov1996,Kazakov1996a,Kazakov2022}).

We start by recalling the classic approach to map enumeration initiated by Tutte in the sixties \cite{Tutte_Census_1962,Tutte_Census_1963,Tutte_enumeration_1968} and which is at the heart of the developments in the EDT (see \cite{Ambjorn_Quantum_1997} for an overview and \cite{Eynard2016,Ambjorn2022} for more recent accounts).

\subsection{Disk function}

The central idea is that, while it is difficult to write an equation for the partition function itself, it is straightforward to obtain one for the generating function of maps with a boundary of controlled length.
Here by \emph{boundary} or \emph{root face} (denoted $f_{\mathrm{r}}$) we simply mean the face that lies on the left of the root edge, and its degree is referred to as the \emph{perimeter} of the map.
One thus considers the \emph{disk generating function}
\begin{align}\label{eq:diskdefinition}
    W^{(\ell)}(t,\mathbf{q}) \coloneqq \sum_{\substack{\text{rooted planar maps }\map\\\deg(f_{\mathrm{r}})=\ell}} \mkern-20mu t^{|\vertices(\map)|} w_\mathbf{q}(\map), \qquad w_\mathbf{q}(\map)\coloneqq\prod_{f \in \faces(\map)\setminus\{f_{\mathrm{r}}\}} q_{\deg f},
\end{align}  
where by convention we set $W^{(0)}(t,\mathbf{q}) = t$, counting the map consisting of a single vertex and no edges.
In the following we will drop the explicit dependence on $\mathbf{q}$ for notational simplicity and simply write $W^{(\ell)}$.

\begin{figure}[h]
    \sidecaption[t]
    \includegraphics[width=.6\linewidth]{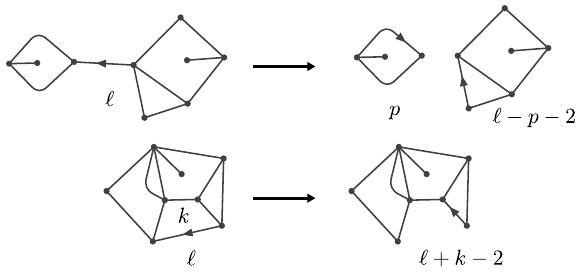}
    \caption{Dropping the root edge of a map of perimeter $\ell$ leads either to a pair of maps of perimeters $p$ and $\ell-p-2$ for some $0\leq p \leq \ell-2$, or to a single map of perimeter $\ell+k-2$ for $k\geq 1$.\label{fig:tutte}}
\end{figure}

For any $\ell \geq 1$ one can decompose a rooted map by removing the root edge, which either leads to a map with one face less or to a pair of maps, see Fig.~\ref{fig:tutte}.
At the level of generating functions this decomposition leads to the famous \emph{Tutte equation} or \emph{loop equation} \cite{Tutte_enumeration_1968}
\begin{equation}\label{eq:tutte}
    W^{(\ell)} = \sum_{k=1}^\infty q_k W^{(\ell+k-2)}+ \sum_{p=0}^{\ell-2} W^{(p)} W^{(\ell-p-2)}. \qquad(\ell \geq 1)
\end{equation}
By introducing a generating variable $x$ for the boundary length $\ell$, 
\begin{equation}\label{eq:diskandpotential}
    W(x) \coloneqq \sum_{\ell=0}^\infty W^{(\ell)} x^{-\ell -1}, \qquad V'(x) \coloneqq x - \sum_{k=1}^\infty q_k x^{k-1},
\end{equation}
this equation can be seen to be equivalent to $W(x)^2 - V'(x) W(x)$ being analytic at $x=0$. 
More precisely, we have the identity
\begin{equation}
    V'(x) W(x) - W(x)^2 = Q(x) \coloneqq t - \sum_{p=0}^\infty x^p \sum_{k=p+2}^\infty q_k W^{(k-p-2)},
\end{equation}
which is solved by
\begin{equation}
    W(x) = \frac{1}{2}\left(V'(x) - \sqrt{V'(x)^2 - 4 Q(x)}\right).
\end{equation}
Here the sign in front of the square root is determined by the requirement that $x W(x) \to t$ as $x\to \infty$.

Let us for the moment assume that only finitely many weights are nonzero, such that $V'(x)$ and $Q(x)$ are both polynomials.
The results, however, can be shown to hold more generally, as we will see in Section~\ref{sec:treebijections}, where we will be more careful about convergence issues.
In the polynomial case one can make the \emph{one-cut assumption} (or use Brown's Theorem \cite{Brown_existence_1965}) that the polynomial $V'(x)^2 - 4 Q(x)$ factorizes as
\begin{equation}
    V'(x)^2 - 4 Q(x) = M(x)^2 (x-c_+)(x-c_-), \qquad (c_+ > c_-)
\end{equation}
where the sign of $M(X)$ is chosen such that $V'(x) / (x M(x)) \to 1$ as $x\to\infty$.
Hence,
\begin{equation}\label{eq:onecutdisksolution}
    W(x) = \frac{1}{2}\left(V'(x) - M(x)\sqrt{(x-c_+)(x-c_-)}\right).\qquad( x\in \mathbb{C} \setminus [c_-,c_+])
\end{equation}
The polynomial $M(x)$ as well as the endpoints $c_\pm = c_\pm(t,\mathbf{q})$ of the branch cut are then completely determined in terms of the weights $t$ and $\mathbf{q}$ by expanding the right-hand side around $x=\infty$ and imposing the condition $x W(x) \to t$ as $x\to \infty$.

This can be made more explicit by performing the Zhukovsky transformation \cite{Eynard2016}
\begin{equation}
    x(z) = \frac{c_++c_-}{2} + \frac{c_+-c_-}{4}\left(z+\frac{1}{z}\right),
\end{equation}
which is designed such that $W(x(z))$ becomes a Laurent polynomial in $z$ (i.e.\ a polynomial in $z$ and $1/z$),
\begin{equation}\label{eq:Wzhukovsky}
    W(x(z)) = \frac{1}{2}V'(x(z)) - M(x(z))\frac{c_+-c_-}{8}\left(z - \frac{1}{z}\right) \equiv \frac{1}{2}V'(x(z)) + y(z).
\end{equation}
The pair of functions $x(z),y(z)$ is known as the \emph{spectral curve} of the model and plays an important role in \emph{topological recursion} \cite{Eynard_Topological_2005,Eynard_Invariants_2007,Eynard2016}, which relates generating functions of maps with multiple boundaries or higher genus to the disk function.
Since we will stick to the planar case, we will not delve into this topic. 

Note that under the transformation $z\to 1/z$ the first term in \eqref{eq:Wzhukovsky} is symmetric, $V'(x(1/z)) = V'(x(z))$, while the second is antisymmetric, $y(z) = - y(1/z)$.
Since $W(x(z)) = t\frac{4}{c_+-c_-} z^{-1} + O(z^{-2})$, it follows that for any $p\geq 0$ we have 
\begin{equation}
    0 = \frac{1}{2}[z^p]V'(x(z)) + [z^p]y(z) = \frac{1}{2}[z^{-p}]V'(x(z)) - [z^{-p}]y(z),
\end{equation}
where the notation $[z^k]f(z)$ for a Laurent polynomial $f$ refers to the coefficient of $z^k$ in $f$.
Hence
\begin{equation}\label{eq:Wtranscoeff}
    [z^{-p}]W(x(z)) = [z^p]V'(x(z)) \stackrel{\eqref{eq:diskandpotential}}{=} \frac{c_+-c_-}{4}\ind_{\{p=1\}} - \sum_{k=1}^\infty q_{k} [z^p] x(z)^{k-1}.
\end{equation}
The equations $[z^0]W(x(z))=0$ and $[z^{-1}]W(x(z)) = 4t/(c_+-c_-)$ then uniquely determine $c_\pm$ in terms of $\mathbf{q}$ and $t$.

The partition function \eqref{eq:boltzpartitionfunction} can be retrieved from $W^{(2)}$ by the observation that zipping open the root edge of a rooted map results bijectively in a rooted map with boundary of length $2$ and at least two faces.
Hence 
\begin{equation}
    Z(t,\mathbf{q}) = W^{(2)}-t^2 \stackrel{\eqref{eq:tutte}}{=} \sum_{k=1}^\infty q_k W^{(k)}.
\end{equation}

\subsection{Pointed maps}

The expression \eqref{eq:onecutdisksolution} for the disk function already displays a degree of \emph{universality}, in that the general structure is independent of the weights $\mathbf{q}$ and $t$. 
This universality becomes more explicit when one considers planar maps with a distinguished face of specified degree \cite{Ambjoern_Multiloop_1990,Ambjorn_Properties_1990,Ambjorn_Quantum_1997,Eynard_Topological_2005}, whose generating functions depend only on $c_\pm$.
Let us concentrate on the special case of planar maps with a distinguished vertex, which are also called \emph{pointed planar maps}.
The generating function $W_\bullet^{(\ell)}(t,\mathbf{q})$ is defined just like $W^{(\ell)}(t,\mathbf{q})$ in \eqref{eq:diskdefinition}, except the sum runs over pointed planar maps and the distinguished vertex does not receive weight $t$.
It should be clear that pointed and unpointed disk functions are related by a $t$-derivative,
\begin{equation}
    W_\bullet^{(\ell)} = \frac{\partial}{\partial t} W^{(\ell)}, \qquad W_{\bullet}(x) = \sum_{\ell=0}^\infty W_\bullet^{(\ell)} x^{-\ell -1} = \frac{\partial}{\partial t} W(x).
\end{equation} 

Inserting \eqref{eq:Wzhukovsky} and applying the chain rule, while observing that $V'(x)$ does not depend on the weight $t$, one finds the relation \cite{Eynard2016}
\begin{align}
    W_{\bullet}(x(z)) = \frac{\partial y(z)}{\partial t} - \frac{y'(z)}{x'(z)} \frac{\partial x(z)}{\partial t}.
\end{align}
Now one should observe that the right-hand side of
\begin{align}
    z x'(z) W_{\bullet}(x(z)) =  z x'(z)\frac{\partial y(z)}{\partial t} - z y'(z)\frac{\partial x(z)}{\partial t}
\end{align}
is a Laurent polynomial that is symmetric under $z \to 1/z$, while the left-hand side approaches $1$ when $z\to \infty$, because $x W_\bullet(x) \to 1$ as $x\to \infty$.
Hence, the full Laurent polynomial must be identically equal to $1$.
Inverting the Zhukovsky transformation then leads to the universal formula
\begin{equation}
    W_\bullet(x) = \frac{1}{\sqrt{(x-c_+)(x-c_-)}}.
\end{equation}

\subsection{Bipartite maps}
If only $q_2,q_4,\ldots$ are nonzero, we are dealing with bipartite planar maps and the branch cut $[c_-,c_+]$ becomes symmetric around $0$. For future reasons we introduce the notation 
\begin{equation}
    R(t,\mathbf{q}) = \frac{1}{4}c_+^2 = \frac{1}{4}c_-^2 = t + O(t^2),
\end{equation}
such that
\begin{equation}
    x(z) = \sqrt{R}\left( z+ \frac{1}{z}\right), \qquad W_\bullet(x) = \frac{1}{\sqrt{x^2 - 4R}}.
\end{equation}
From the last formula we deduce by series expansion around $x=\infty$ that
\begin{equation}\label{eq:pointedevendisk}
    W_\bullet^{(2\ell)} = \binom{2\ell}{\ell} R^\ell.
\end{equation}
Note in particular that $W_\bullet^{(2)} = 2R$.
By removing the contribution $2t$ of the maps consisting of a single edge and zipping closed the boundary of the remaining maps, we obtain the generating function 
\begin{equation}\label{eq:pointedZ}
    \frac{\partial}{\partial t}Z = 2R - 2t
\end{equation}
for pointed bipartite planar maps.
Equation \eqref{eq:Wtranscoeff} with $p=1$ and $[z^{-1}]W(x(z)) = t/\sqrt{R}$ results in the explicit recursive equation
\begin{equation}\label{eq:recursive_eq_R}
    R = t + \sum_{k=1}^\infty q_{2k} \binom{2k-1}{k} R^{k}.
\end{equation}
The simple form of this equation and the universal form of the pointed disk function $W_\bullet(z)$ have appeared rather miraculously.
In the following sections we will give two explanations for this simplicity, a bijective approach involving combinatorial trees and a probabilistic approach involving a peeling exploration.
As a bonus, both approaches provide insights into the geometry of the $\mathbf{q}$-Boltzmann maps.

\section{Bijection with trees}\label{sec:treebijections}

The solution method presented to determine the generating function of planar maps required some ingenuity (that can be traced back to Tutte): the generating function $Z(t,\mathbf{q})$ could not be identified as a solution of an equation, but introducing an additional generating variable $x$ for the root face degree such an equation could be found, which miraculously could be solved rather explicitly.
Compared to planar maps, trees are much simpler objects because their generating functions do naturally satisfy an equation (without introducing extra variables) and therefore feature prominently in the combinatorial literature (see \cite{Flajolet2009} for an overview).
A natural strategy to enumerate non-treelike objects, or to explain a mysterious simplicity in the enumeration, is to seek bijective relations with trees.
This route has played a central role in the mathematical developments of planar maps and a good number of examples of such tree bijection are known, see    
\cite{Cori_Planar_1981,Schaeffer1997,Schaeffer_Conjugaison_1998,BDFG04,Poulalhon2006,Bernardi2012} for a nonexhaustive list.
We will focus on the Bouttier--DiFrancesco--Guitter bijection \cite{BDFG04} that is well suited for the enumeration of planar maps with control on the face degrees.
We restrict our attention to bipartite planar maps and refer the reader to \cite{BDFG04} for the general case.

\subsection{The Bouttier--DiFrancesco--Guitter bijection}

\begin{figure}[t]
    \centering
    \includegraphics[width=.75\linewidth]{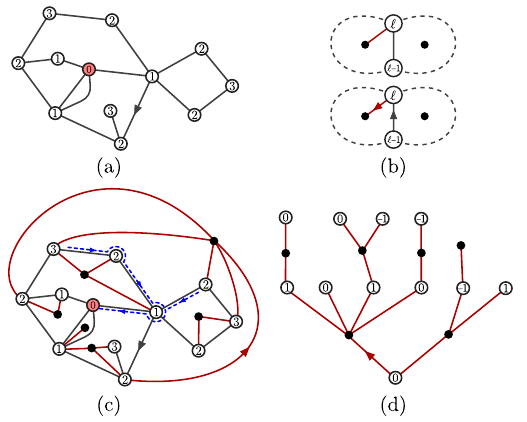}
    \caption{The Bouttier--DiFrancesco--Guitter bijection. (a) A rooted even planar map $\map$ with a distinguished vertex (shaded in red) together with its canonical labeling by the graph distance. (b) The prescription for drawing new (red) edges and a new root. (c) The result of applying the prescription to $\map$. The (blue) dashed lines indicate two left-most geodesics. (d) After deleting the edges of $\map$ and the origin and shifting the labels such that the start of the root edge receives label $0$, one obtains the mobile $\mathfrak{t}$.}
    \label{fig:bdfg}       
\end{figure}

In the last section we have seen that particularly pointed maps admit simple generating functions, so let us consider a rooted bipartite planar map $\map$ with a distinguished vertex, that we call the \emph{origin}. 
Naturally one may assign a label $\ell_v$ to each vertex $v \in \vertices(\map)$ by taking $\ell_v$ to be the graph distance along the edges of $\map$ from $v$ to the origin (Fig.~\ref{fig:bdfg}a).
Because $\map$ is bipartite, the labels at the endpoints of each edge differ exactly by $1$.
For reasons that will become clear soon, let us restrict to the situation where the labels along the root edge of $\map$ increase from its start to end.
This is the case for exactly half of the maps, so by \eqref{eq:pointedZ} these should be enumerated by $R - t$, for which we will deduce a bijective explanation.

The \emph{Bouttier--DiFrancesco--Guitter bijection} (BDFG) provides an encoding of these maps in terms of so-called mobiles. 
A \emph{mobile} is a tree $\mathfrak{t}$, i.e.\ a rooted planar map with only one face, with black vertices and integer-labeled white vertices satisfying the following properties (see Fig.~\ref{fig:bdfg}d for an example):
\begin{enumerate}[label = (\roman*)]
    \item The two endpoints of each edge have different color.\label{item:mobile1}
    \item The root edge starts at a white vertex with label $0$.\label{item:mobile2}
    \item Around each black vertex, if a white neighbour has label $\ell$ then the next white neighbour in clockwise order around the black vertex must have label at least $\ell-1$.
\end{enumerate}

\subsection{From planar maps to trees}\label{sec:mapstotrees}

The procedure to associate a mobile $\mathfrak{t}$ to $\map$ is as follows:
\begin{enumerate}
    \item Let the vertices of $\map$ be white and draw a new black vertex in each face of $\map$ (Fig.~\ref{fig:bdfg}c).
    \item For each edge $e$ of $\map$, let $v$ be the endpoint of $e$ with the largest label. Draw a new edge starting at $v$ and ending on the black vertex within the face to the left of $e$ when facing $v$ (Fig.~\ref{fig:bdfg}b). If $e$ is the root of $\map$ we take the new edge to be the new root (oriented away from $v$).
    \item Remove all original edges of $\map$ as well as the origin vertex (Fig.~\ref{fig:bdfg}d). 
    \item Shift all labels uniformly such that the root vertex receives label $0$.
\end{enumerate}
Why does this procedure result in a mobile? 
The main feature of the construction is that $\mathfrak{t}$ cannot have any cycles.
The explanation is that for any edge $e$ of $\map$ one can find a curve, called the \emph{left-most geodesic}, starting at (say, the midpoint of) $e$ and ending at the origin that does not intersect $\mathfrak{t}$.
This immediately implies the impossibility of cycles in $\mathfrak{t}$, because every cycle in $\mathfrak{t}$ would enclose the origin on one side and at least one edge on the other side, contradicting the existence of a left-most geodesic path from that edge to the origin.

The left-most geodesic is constructed as follows (see the dashed curve in Fig.~\ref{fig:bdfg}c for an example): denote the endpoints of $e$ by $v_\ell$ and $v_{\ell-1}$ with labels $\ell$ and $\ell-1$ respectively. 
The curve starts by traversing $e$ towards $v_{\ell-1}$.
If $\ell=1$, $v_{\ell-1}$ is the origin and we are done. 
Otherwise, the curve circles around $v_{\ell-1}$ in clockwise direction until it encounters an edge with endpoint at distance $\ell-2$, that we denote $v_{\ell-2}$.
Such an edge always exists due to the definition of the graph distance, and by construction of $\mathfrak{t}$ one encounters no edge of $\mathfrak{t}$ along the way.
Traversing the edge to $v_{\ell-2}$ and iterating, one obtains a curve ending at the origin $v_0$, since that is the unique vertex with minimal label.
The path $v_\ell, v_{\ell-1}, \ldots , v_0$ in $\map$ is called the left-most geodesic, because it is a path of minimal length from $v_\ell$ to the origin and at each vertex it chooses the left-most option among such minimal paths. 

In the absence of cycles, the number of connected components of $\mathfrak{t}$ is given by $|\vertices(\mathfrak{t})| - |\edges(\mathfrak{t})|$.
But by construction $|\vertices(\mathfrak{t})| = |\vertices(\map)| + |\faces(\map)| - 1$ and $|\edges(\mathfrak{t})| = |\edges(\map)|$, which together with Euler's formula \eqref{eq:eulerformula} implies that $\mathfrak{t}$ has a single connected component and is thus a tree.
That the labels satisfy the properties of a mobile is straightforwardly checked from the construction.

\subsection{From trees to planar maps}\label{sec:treetomap}
Starting from a mobile $\mathfrak{t}$ one constructs a map $\map$ in a reverse fashion. The angular region around a vertex $v$ that is delimited by two neighbouring edges incident to $v$ is called a \emph{corner} of $v$. 
The \emph{contour} of a face $f$ is the cyclic sequence of corners one encounters while walking around the perimeter of $f$ while keeping the edges on the right-hand side.
The contour of (the unique face of) a tree thus visits all its corners in clockwise direction. 
The procedure is then as follows:
\begin{enumerate}
    \item Add a new white vertex (the origin) with label $\ell_{\mathrm{min}}-1$ in the face of $\mathfrak{t}$, where $\ell_{\mathrm{min}}$ is the minimal label of $\mathfrak{t}$.
    \item For each corner $c$ of a white vertex with label $\ell$ in $\mathfrak{t}$, we draw a new edge from $c$ to the next corner of a white vertex in the contour that has label $\ell - 1$ in case $\ell > \ell_{\mathrm{min}}$ or to the origin in case $\ell=\ell_{\mathrm{min}}$. If $c$ is the corner of the root vertex that sits left of the root edge, then the new edge is taken to be the new root (oriented away from $c$).
    \item Remove all edges of $\mathfrak{t}$.
\end{enumerate}
One can show \cite{BDFG04} that the construction is well-defined for any mobile, in the sense that the edges in the second step can be drawn unambiguously in a non-intersecting fashion, and that this is precisely the inverse of the construction in Section~\ref{sec:mapstotrees}.
Note that the labels of $\mathfrak{t}$ need to be shifted by  $1-\ell_{\mathrm{min}}$ to arrive at the graph distances to the origin.

Let us make an observation about the left-most geodesics that will become important later \cite{LeGall_topological_2007}.
We have precisely one such geodesic $v_{\ell},v_{\ell-1},\ldots,v_{\ell_{\mathrm{min}}-1}$ of length $\ell - \ell_{\mathrm{min}}+1$ for each corner $c$ of a white vertex with label $\ell$ in $\mathfrak{t}$ and the path can be easily deduced from the sequence of labels in the contour of $\mathfrak{t}$: for $\ell_{\mathrm{min}} \leq i < \ell$ the vertex $v_i$ is simply the first vertex with label $i$ encountered when following the contour starting from $c$.
In particular, two geodesics from corner $c$ at vertex $v$ and coner $c'$ at vertex $v'$ will typically merge before reaching the origin (see Fig.~\ref{fig:bdfg}c). 
The merge happens at a vertex with label $\max(k,k')-1$ where $k$ is the minimal label along the contour between $c$ (inclusive) and $c'$ (exclusive), and $k'$ is the minimal labels in the contour between $c'$ (inclusive) and $c$ (exclusive).
Although one cannot easily deduce the graph distance $d_{\mathrm{gr}}(v,v')$, we do find an upper bound by concatenating the geodesics up to their merger,
\begin{equation}\label{eq:dgrinequality}
    d_{\mathrm{gr}}(v,v') \leq \ell_v + \ell_{v'} - 2\max(k,k')+2.
\end{equation}

\subsection{Enumeration based on the trees}\label{sec:treeenumeration}



In the BDFG bijection each face of $\map$ of degree $2k$ corresponds to a black vertex in $\mathfrak{t}$ of degree $k$ and each vertex, except for the distinguished one, to a white vertex of $\mathfrak{t}$.
As a consequence, the generating function $R(t,\mathbf{q}) - t$ from \eqref{eq:pointedZ} for (half of) the pointed bipartite planar maps is also the generating function of mobiles with at least one edge and a weight $t$ per white vertex and a weight $q_{2k}$ per black vertex of degree $k$. 
These mobiles admit a convenient recursive decomposition.
Let us denote the root vertex by $v_0$.
If the degree of the black vertex at the end of the root edge is $k$, then it has $k-1$ white children $v_1,\ldots,v_{k-1}$. 
Each vertex $v_i$ together with its offspring, excluding the branch of the root edge in case of $v_0$, determines a mobile, once the labels have been shifted such that the root vertex $v_i$ receives label $0$.
Noting that these mobiles may take the form of a single white vertex with no children, this leads immediately to the equation
\begin{equation}\label{eq:Rrecursionillustrated}
    R - t = \sum_{k=1}^\infty q_{2k} \sum_{\substack{\text{labels on}\\v_1,\ldots,v_{k-1}}}\vcenter{\hbox{\includegraphics[width=1.7cm]{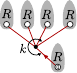}}} = \sum_{k=1}^\infty q_{2k} \binom{2k-1}{k} R^{k},
\end{equation}
because there are precisely $\binom{2k-1}{k}$ choices for the labels $\ell_1,\cdots, \ell_{k-1}$ on $v_1,\ldots,v_{k-1}$ satisfying the requirements $\ell_1 \geq -1$, $\ell_k \leq 1$, and $\ell_{i+1} \geq \ell_i-1$ for $i=1, \ldots, k-1$.
We thus reproduce equation \eqref{eq:recursive_eq_R}.
Since the summand gives the contribution of maps with root face of degree $2k$, we also reproduce
\begin{equation}
    W_\bullet^{(2k)} = 2 \binom{2k-1}{k} R^k = \binom{2k}{k} R^k,
\end{equation}
where the factor of $2$ compensates for the fact that only half of the edges adjacent to the root face have increasing label and can thus serve as root edge.

\subsection{Admissibility and criticality}
\label{sec:admissibilitycriticality}

So far we have ignored issues of convergence in the computations of partition functions, which is ok if one chooses to work only at the level of formal generating series.
However, soon we will be questioning the statistical properties of random maps, so we better make sure that the probability measures are sane.
Luckily the tree bijections allow one to easily deduce criteria on the weight sequence $\mathbf{q}$.
Restricting to non-pathological cases where at least one of $q_4,q_6,\ldots$ is non-zero, we say $\mathbf{q}$ is \emph{admissible} when the generating function of rooted, pointed planar maps (with weight $t=1$ per vertex) is finite, $R(\mathbf{q}) = R(t=1,\mathbf{q}) < \infty$. 
This implies the same for (unpointed) rooted planar maps, $Z(\mathbf{q}) < \infty$ and those with boundary, $W^{(2\ell)}(\mathbf{q}) < \infty$ for all $\ell \geq 1$. 
With some extra work one can show that the converse is true as well \cite{Bernardi2019,Curien2019}, namely that $Z(\mathbf{q}) < \infty$ or $W^{(2\ell)}(\mathbf{q}) < \infty$ for some $\ell \geq 1$ implies $R(\mathbf{q}) < \infty$, so any of these criteria can be used as definition of admissibility.

From equation \eqref{eq:Rrecursionillustrated} it follows that a necessary condition for $\mathbf{q}$ to be admissible is that the equation
\begin{equation}\label{eq:gofR}
   g_\mathbf{q}(r) = 1, \quad\text{where}\quad g_\mathbf{q}(r) = r - \sum_{k=1}^\infty q_{2k}\binom{2k-1}{k} r^k
\end{equation}
has at least one solution, since $g_\mathbf{q}(R) = 1$ when $\mathbf{q}$ is admissible.
It turns out that this is also sufficient and that $R(\mathbf{q})$ is given by the smallest positive fixed point \cite{Marckert2007}. 
The reasoning is instructive, so we will summarize it here.

If $g_\mathbf{q}(r_1)=1$ we have the identity
\begin{equation}
    r_1^{-1} + \sum_{k=1}^\infty q_{2k} \binom{2k-1}{k} r_1^{k-1} = 1.
\end{equation}
Since each term is positive, we may interpret them as probabilities and explicitly construct a random mobile as follows.
We start with a single white node, that we designate to be active. 
Then at each step we visit each active white node and with probability $r_1^{-1}$ we deactivate the node or with probability $\binom{2k-1}{k} r_1^{k-1}$ we insert a black descendant which in turn has $k-1$ new active white descendants.
The crux is to determine whether this random process produces a finite or an infinite tree.
By adding up the probabilities of all finite mobiles thus produced, one finds that the mobile will be finite with probability
\begin{equation}\label{eq:finitemobileprob}
    \frac{1}{r_1} \sum_{\text{mobiles }\mathfrak{t}} \,\,\,\prod_{\text{black vertices }v} \binom{2\deg v-1}{\deg v} q_{2\deg v}.
\end{equation}
On the other hand, the number of active white nodes at each step in our construction has precisely the law of a \emph{Bienayme--Galton--Watson (BGW) process}.
It is well known that the probability of extinction of such a process is $1$ if and only if the mean offspring per individual is less or equal to $1$.
In our case the mean offspring is $\sum_{k=1}^\infty k\,q_{2k} \binom{2k-1}{k} r_1^{k-1} = 1-g_\mathbf{q}'(r_1)$.
Since $g_\mathbf{q}(0) = 0$ and $r_1$ is the first solution to $g_\mathbf{q}(r)=1$, we must have $g_\mathbf{q}'(r_1) \in [0,1)$.
So the mean offspring is at most $1$, implying that the probability \eqref{eq:finitemobileprob} equals $1$ and therefore $R(\mathbf{q}) = r_1$. 
This verifies our claim.

This last observation naturally leads to a distinction between admissible sequences $\mathbf{q}$ that are \emph{subcritical}, if $g_\mathbf{q}'(R) > 0$, and those that are \emph{critical}, if $g_\mathbf{q}'(R) = 0$.
To understand the difference, we can have a look at the generating function $R(t,\mathbf{q})$ that includes a weight $t \in [0,1]$ per vertex, which satisfies 
\begin{equation}\label{eq:Wginv}
    g_\mathbf{q}(R(t,\mathbf{q})) = t.   
\end{equation}
The probability that a (unpointed but rooted) $\mathbf{q}$-Boltzmann map has precisely $n$ vertices is 
\begin{equation}
    \mathbb{P}_{\mathbf{q}}(n\text{ vertices}) = \frac{1}{n Z}[t^{n-1}]R(t,\mathbf{q}) = \frac{1}{n Z}[t^{n-1}]g_\mathbf{q}^{-1}(t).
\end{equation}
For large $n$ this probability is thus determined by singularity analysis of $g_\mathbf{q}^{-1}$.

Let's first focus on the case where $g_\mathbf{q}$ has radius of convergence larger than $R(\mathbf{q})$, in which case $\mathbf{q}$ is called \emph{regular}. 
This happens for instance when only a finite number of weights $q_{2k}$ are non-zero.
In the subcritical case $g_\mathbf{q}'(R) > 0$, we have that $g_\mathbf{q}^{-1}(t)$ has radius of convergence larger than $1$. Therefore the number of vertices has an exponential tail: there exists a $c>0$ such that for all $n\geq 0$,
\begin{equation}
    \mathbb{P}_{\mathbf{q}}(n\text{ vertices}) \leq e^{-c n}.\qquad\text{(regular subcritical)}
\end{equation}
In the critical case $g_\mathbf{q}'(R) = 0$ we necessarily have $g_\mathbf{q}''(R) < 0$ and therefore $g_\mathbf{q}^{-1}(t) = R - \sqrt{\frac{1-t}{\frac{1}{2}|g_\mathbf{q}''(R)|}} + o(\sqrt{1-t})$.
The same expansion applies to the more general case of \emph{generic critical} $\mathbf{q}$, in which $g_\mathbf{q}$ is allowed to have a radius of convergence as small as $R(\mathbf{q})$, but for which  $g_\mathbf{q}''(R)$ is still finite.
Singularity analysis then implies that\footnote{Here and in the following we will use the notation $f(n)\stackrel{n\to\infty}{\sim}g(n)$ if $f(n)$ is asymptotic to $g(n)$ as $n\to\infty$, i.e.\ when $\lim_{n\to\infty} f(n) / g(n) = 1$.} 
\begin{equation}
    \mathbb{P}_{\mathbf{q}}(n\text{ vertices}) \stackrel{n\to\infty}{\sim} c \,n^{\gamma_{\mathrm{s}}-2},\qquad \gamma_{\mathrm{s}} = - \frac{1}{2}.\qquad\text{(generic critical)}
\end{equation}
The \emph{string susceptibility exponent} $\gamma_{\mathrm{s}}$ is thus a universal critical exponent for generic critical Boltzmann maps.

\begin{figure}[t]
    \centering
    \sidecaption[t]
    \includegraphics[width=.6\linewidth]{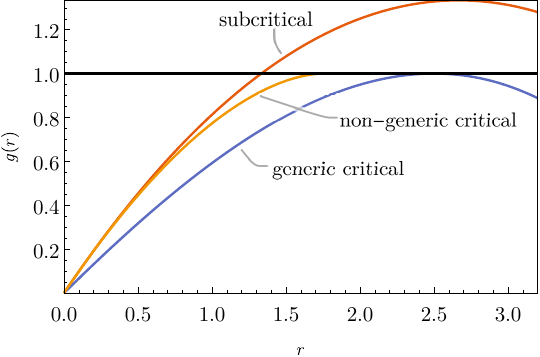}
    \caption{Examples of the curves $g_\mathbf{q}(r)$ for subcritical, regular critical, and non-generic critical weight sequences $\mathbf{q}$.}
    \label{fig:criticalplot}       
\end{figure}

One may escape this universality only when $g_\mathbf{q}$ has radius of convergence exactly equal to $R(\mathbf{q})$ and $g_\mathbf{q}''(R) = -\infty$.
In particular, $\mathbf{q}$ is called \emph{non-generic critical of type $a$} if \cite{LeGall_Scaling_2011,Budd_Geometry_2017,Ambjorn_Generalized_2016,Borot_recursive_2012}
\begin{equation}
    g_\mathbf{q}(r) = 1 - C\,(R-r)^{a-\tfrac12} + o((R-r)^{a-\tfrac12}).
\end{equation}
Observe that we need to take $a \in (3/2,5/2)$ to ensure $g_\mathbf{q}'(R)=0$ and $g_\mathbf{q}''(R) = -\infty$.
In this case $g_\mathbf{q}^{-1}(t) = R - (\frac{1}{C}(1-t))^{2/(2a-1)} + o\left((1-t)^{2/(2a-1)}\right)$ and therefore one finds
\begin{equation}
    \mathbb{P}_{\mathbf{q}}(n\text{ vertices}) \sim c \,n^{\gamma_{\mathrm{s}}-2},\qquad \gamma_{\mathrm{s}} = - \frac{2}{2 a-1}.\qquad\text{(non-generic critical)}
\end{equation}
An example \cite{Ambjorn_Multi_2016,Budd_Geometry_2017} of such a non-generic critical weight sequence $\mathbf{q}$ of type $a\in(3/2,5/2)$ is 
\begin{equation}
    q_{2k} = 2\cos(a\pi)\frac{\Gamma(\tfrac12+a)\Gamma(\tfrac12+k-a)}{\Gamma(\tfrac12)\Gamma(\tfrac12+k)}(4a-2)^{-k} \ind_{\{k\geq 2\}},
\end{equation}
for which
\begin{equation}
    R(\mathbf{q}) = a - 1/2, \qquad g_\mathbf{q}(r) = 1 - (1 - r/R)^{a-\frac12}.
\end{equation}

In full generality it holds that an admissible sequence $\mathbf{q}$ is critical if and only if the number of vertices of a $\mathbf{q}$-Boltzmann planar map has infinite variance.
Critical random maps are therefore much more likely to be very large than subcritical ones, making them the natural choice to investigate scaling limits.
Perhaps more importantly, we will see later in Section~\ref{sec:peelinginfinitemaps} that critical random maps naturally occur within \emph{infinite maps}.

\subsection{Geodesic distance statistics}\label{sec:distancestatistics}

Besides providing a combinatorial interpretation to the simple enumeration formulas for pointed maps, the tree bijection provides a natural way to study geodesic distances.
To illustrate this, let us focus on the simplest example of quadrangulations, i.e.\ $q_{k} = q_4 \delta_{k,4}$, referring the interested reader to \cite{Bouttier_Geodesic_2003} for the general case. 
From the previous discussion it easily follows that random quadrangulations are subcritical when $q_4 < 1/12$ and generic critical for $q_4 = 1/12$.

\begin{figure}[t]
    \centering
    \includegraphics[width=.85\linewidth]{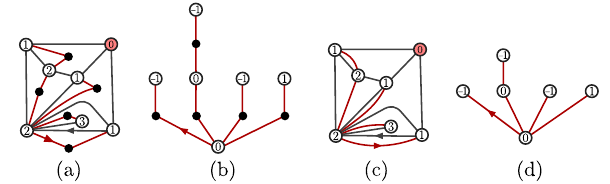}
    \caption{(a) The BDFG bijection applied to the quadrangulation of Fig.~\ref{fig:planarmap} results in the mobile (b) with all black vertices of degree 2. Merging the edges incident to black vertices results in the Cori--Vauqelin--Schaeffer bijection between rooted pointed quadrangulations (c) and labeled plane trees (d).}
    \label{fig:cvsbijection}
\end{figure}

In the corresponding mobiles the black vertices all have degree two, meaning that we may as well merge the pair of edges adjacent to each black vertex to obtain a tree with white vertices only (Fig.~\ref{fig:cvsbijection}). 
The labels between neighboring white vertices are then seen to differ by at most $1$, such that \eqref{eq:Rrecursionillustrated} becomes
\begin{equation}\label{eq:Rquad}
    R - t = 3 q_4 R^2.
\end{equation}
This bijection between pointed quadrangulations and labeled trees is precisely the \emph{Cori--Vauqelin--Schaeffer bijection} \cite{Cori_Planar_1981,Schaeffer_Conjugaison_1998}.
One can approach the enumeration of quadrangulations with control on geodesic distances \cite{Bouttier_Geodesic_2003} by introducing the generating function $R_d$ of quadrangulations in which the labels increase along the root edge and such that the start of the root edge has label at most $d$.
Then $R_0 = Z$, because for $d=0$ the origin must be located at the start of the root edge, and $\lim_{d\to\infty} R_d = R$, while 
\begin{equation}\label{eq:quadtwopoint}
    G_d=R_{d}-R_{d-1}
\end{equation} 
is the generating function of pointed quadrangulations in which the start and end of the root edge are at distance exactly $d$ and $d+1$ from the origin, respectively.
The reason to work with $R_d$ instead of directly with $G_d$ is that $R_d$ satisfies a recursion equation that generalizes \eqref{eq:Rquad},
\begin{equation}\label{eq:Rnrecurrence}
    R_d - t = q_4 R_d(R_{d-1}+R_d+R_{d+1}).
\end{equation}
We can understand this equation by interpreting $R_d$ as the generating function of trees with positive integer labels that differ by at most $1$ between neighbors and such that the root vertex receives label $d+1$.
Observing that the endpoint of the root edge has label $d$, $d+1$, or $d+2$ leads to the above formula.

Equation \eqref{eq:Rnrecurrence} can be solved explicitly \cite{Bouttier_Geodesic_2003}, yielding
\begin{equation}\label{eq:Rdsolution}
    R_d = R \frac{(1-x^{d+1})(1-x^{d+4})}{(1-x^{d+2})(1-x^{d+3})},\qquad x+ \frac{1}{x}+4 = \frac{1}{q_4 R}, \quad |x|<1.
\end{equation}

This exact expression allows us already to deduce some statistics concerning geodesic distances in large random quadrangulations.
For instance, we can consider the situation where we condition a pointed critical quadrangulation $\map$ to have its distance $d_\bullet(\map)$ between the root vertex and the origin to be exactly equal to $d$ and then ask about the distribution of the size of $\map$.
It satisfies 
\begin{align*}
    \mathbb{E}\left[e^{-\lambda |\faces(\map)|} \middle| d_\bullet(\map) = d\right] = \frac{G_d(q_4=\tfrac{1}{12}e^{-\lambda})}{G_d(q_4=\tfrac{1}{12})}.
\end{align*}
Inserting \eqref{eq:quadtwopoint} and \eqref{eq:Rdsolution}, one may check that to achieve a non-trivial limit one should scale $\lambda$ proportionally to $d^{-4}$ as $d\to\infty$.

Put differently, if we let $\lambda = \Lambda \epsilon^2$ for $\Lambda>0$ fixed and consider $|\faces(\map)| \epsilon^2$ to be a rescaled area of $\map$, then 
$R(\tfrac{1}{12}e^{-\Lambda \epsilon^2}) = 2-2\sqrt{\Lambda} \epsilon + O( \epsilon^2)$ and $x(\tfrac{1}{12}e^{-\Lambda \epsilon^2})=1-\sqrt{6}\sqrt[4]{\Lambda} \sqrt{\epsilon} + O(\epsilon)$.
Keeping $d \sqrt{3\epsilon /2} = D$ fixed while sending $d\to\infty$, leads to a non-trivial limit $x^d = e^{-2\sqrt[4]{\Lambda}D} + O(\sqrt{\epsilon})$.
Hence 
\begin{equation}\label{eq:geodtwopoint}
    \lim_{\epsilon\searrow 0}\mathbb{E}\left[e^{-\Lambda |\faces(\map)| \epsilon^2} \middle| d_\bullet(\map) = \lfloor \sqrt{2/3} \,D / \sqrt{\epsilon}\rfloor \right] = \Lambda^{3/4}D^3 \frac{\cosh \sqrt[4]{\Lambda} D}{\sinh^3 \sqrt[4]{\Lambda} D}.
\end{equation}
The right-hand side is known as the \emph{geodesic two-point function} of two-dimensional quantum gravity with cosmological constant $\Lambda$, which was identified first by Ambjorn and Watabiki in \cite{Ambjorn_Scaling_1995}.
In particular,
\begin{align*}
  \lim_{\epsilon\searrow 0}\mathbb{E}\left[|\faces(\map)| \,\middle| d_\bullet(\map) = \lfloor \sqrt{2/3} \,D / \sqrt{\epsilon}\rfloor \right] = \frac{D^4}{15},
\end{align*}
which signals that typical geodesic distances in a quadrangulation with $n$ faces are of order $n^{1/4}$.

\section{Continuum limit: Brownian geometry}\label{sec:continuumlimit}

The previous calculation has shown that the distribution of the geodesic distance between a single pair of random points (in this case the root and origin vertex) is under analytic control and that one can study its scaling limit.
It is natural to ask whether this can be generalized to the distances between \emph{all} pairs of points simultaneously for random maps of increasing size.
This is a question about scaling limits of metric spaces. 

\subsection{From maps to metric spaces}

Recall that a \emph{metric space} is a pair $(V,d)$ consisting of a set $V$ and a distance function $d : V \times V \to \mathbb{R}$ satisfying $d(x,x) = 0$ for $x\in V$, $d(x,y)=d(y,x)>0$ when $x\neq y$ and the triangle inequality $d(x,z) \leq d(x,y)+d(y,z)$ for all $x,y,z\in V$.
We denote the space of all compact metric spaces, viewed up to isometry, by $\mathbb{M}$.
Note that this is a huge space: it contains all finite metric spaces, all metric spaces induced by compact Riemannian manifolds of arbitrary dimension and topology, but also much wilder spaces.

There are multiple ways one can associate a metric space to a map $\map$, but a practical choice in light of the previous bijection is to consider the finite metric space $(\vertices(\map),d_{\mathrm{gr}})\in\mathbb{M}$, i.e.\ the set of vertices equipped with the graph distance. 
Alternatives are the dual graph distance on the set of faces, that we will encounter in Section~\ref{sec:peelinggeometry}, or the Riemannian metric space induced by the gluing of regular polygons (as studied for example in \cite{Carrance_Convergence_2021}).
For maps with not too large face degrees, in particular generic critical Boltzmann maps, one expects this choice to have little influence on scaling limits. 
For non-generic critical maps the situation is different, as we will see in Section~\ref{sec:peeling}

If we take $\map$ to be a uniform quadrangulation with $n$ faces, then our previous discussion suggests that the metric space $(\map,n^{-1/4}d_{\mathrm{gr}})$, in which the graph distance is normalized by the typical distance $n^{1/4}$ between random vertices, somehow approaches a continuous random metric space.
More generally we could consider a $\mathbf{q}$-Boltzmann planar map and condition on the number $n$ of faces.

\subsection{The Gromov--Hausdorff topology}

What does it mean for a sequence of random metric spaces in $\mathbb{M}$ to have a limit? In order to make sense of this, we need to be able to quantify similarity between metric spaces. 
This is achieved by the \emph{Gromov--Hausdorff distance} $d_{\mathrm{GH}}$ on $\mathbb{M}$ .
We will not provide a full definition but provide an equivalent characterization in terms of correspondences \cite[Sec.~7.3.3]{Burago_course_2001}. 
A \emph{correspondence} between sets $V_1$ and $V_2$ is a subset $R \subset V_1 \times V_2$ such that each element of $V_1$ and each element of $V_2$ occurs at least once in a pair in $R$.
We should thus think of a correspondence as a many-to-many mapping between $V_1$ and $V_2$.
If $V_1$ and $V_2$ are metric spaces, with distances $d_1$ and $d_2$ respectively, then the \emph{distortion}
\begin{equation}
    \operatorname{dis}(R) = \sup_{(x_1,x_2),(y_1,y_2)\in R} \left|d_1(x_1,y_1) - d_2(x_2,y_2)\right| 
\end{equation}
of a correspondence $R$ quantifies how far this mapping is from being an isometry.
The Gromov--Hausdorff distance $d_{\mathrm{GH}}$ between metric spaces $(V_1,d_1)$ and $(V_2,d_2)$ is then (half) the minimal distortion possible,
\begin{equation}
    d_{\mathrm{GH}} \left( (V_1,d_1), (V_2,d_2) \right) = \frac{1}{2} \inf_{\text{correspondences }R} \operatorname{dis}(R).
\end{equation}
Remarkably the Gromov--Hausdorff distance turns the space $\mathbb{M}$ of all compact metric spaces into a metric space itself, with several pleasant properties like being complete (every Cauchy sequence has a limit) and separable (it contains a countable dense subset). 
A \emph{random metric space} is nothing but a probability measure on $\mathbb{M}$.
The Gromov--Hausdorff distance or, more precisely, the topology it induces on $\mathbb{M}$ allows one to decide whether a sequence of random metric spaces converges in distribution to a limiting random metric space.

\begin{figure}[t]
    \centering
    \includegraphics[width=\linewidth]{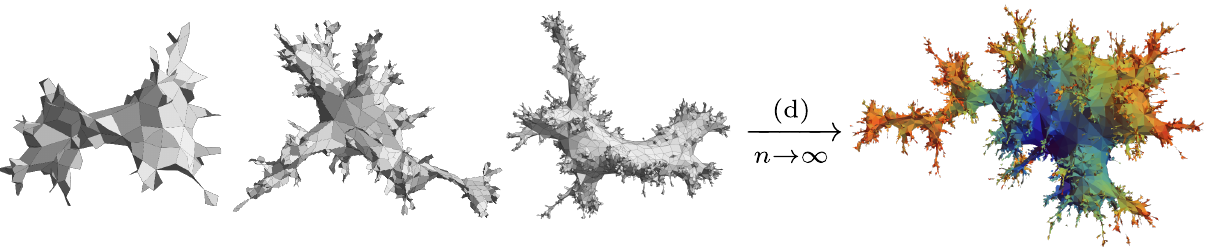}
    \caption{Visualizations of simulated uniform quadrangulations with increasing number $n$ of faces as well as the Brownian sphere (on the right). The images are an attempt at faithfully embedding the metric spaces in 3D Euclidean space.}
    \label{fig:brownianmap}
\end{figure}

Now we are in a position to formulate precise scaling limit results.
If $\mathbf{q}$ is generic critical, and $\map_n$ is a $\mathbf{q}$-Boltzmann planar map conditioned to have $n$ vertices (or $n$ faces or $n$ edges) then there exists a constant $C > 0$ such that we have the convergence in distribution in the Gromov--Hausdorff topology 
\begin{equation}\label{eq:browniansphere}
    (\map_n, C n^{-1/4}d_{\mathrm{gr}}) \xrightarrow[n\to\infty]{(\mathrm{d})} \mathbf{B}
\end{equation}
towards a random metric space $\mathbf{B}$ that is called the \emph{Brownian map} or \emph{Brownian sphere} (see Fig.~\ref{fig:brownianmap}).
This famous result was first proved by Le~Gall \cite{LeGall2013}, in the case of regular critical $\mathbf{q}$-Boltzmann maps (as well as uniform triangulations) when conditioned on the number of faces, and simultaneously using different methods by Miermont \cite{Miermont_Brownian_2013}, in the case of quadrangulations.
The extension to the generic critical case and conditioning on any of the vertices, edges and faces, is due to Marzouk \cite{Marzouk2018}.
Many other families of maps have been shown to share the same limit, like uniform maps \cite{Bettinelli_scaling_2014}, uniform simple triangulations and quadrangulations \cite{Addario-Berry_scaling_2017}, non-bipartite Boltzmann maps \cite{Addario-Berry_Convergence_2021}, uniform cubic planar graphs \cite{Albenque_Random_2022}, and more.
In the case of uniform triangulations, it is also known that the convergence is robust under local deformations of the metric \cite{Curien_First_2019}.

Discussing the full proof of \eqref{eq:browniansphere} is beyond the scope of this chapter, but we can highlight the important ingredients. 
Importantly, we need to understand the Brownian sphere and it should not come as a surprise that its construction heavily relies on random trees.

\subsection{Continuum Random Tree}\label{sec:crt}

Recall that a pointed rooted quadrangulation with $n$ faces, with the extra condition that the distance from the origin increases along the root edge, is uniquely encoded by a labeled rooted plane tree with $n$ edges. 
The labels are allowed to differ by at most $1$ along the edges.
This means that each rooted plane tree admits precisely $3^n$ different labelings.
In particular, if the quadrangulation is chosen uniformly at random, then the associated tree, after forgetting its labels, is a uniform random plane tree with $n$ edges.
Seen as metric spaces, when equipped with the graph distance, such trees admit a well-known scaling limit themselves: the \emph{continuum random tree (CRT)} introduced by Aldous \cite{Aldous1991}.
One can view this statement as a limit in the Gromov--Hausdorff sense, analogously to \eqref{eq:browniansphere}, but with a normalization $n^{-1/2}$ instead of $n^{-1/4}$.

However, in the case of plane trees there is a stronger topology that is also easier to work with, namely convergence at the level of contour functions.
Recall the definition of the contour of a plane tree $\mathfrak{t}$ from Section~\ref{sec:treetomap}. 
Let $C_\tree(i)$ for $i=0,1,\ldots,2n$ be the graph distance from the $i$th corner in the contour to the root vertex.
By linear interpolation this gives rise to the \emph{contour function} $C_\tree : [0,2n] \to \mathbb{R}_{\geq 0}$.
In the case of a uniform plane tree, $C_\tree$ has the law of a random walk with increments $\pm 1$ started at $C_\tree(0)=0$ and conditioned to stay non-negative before returning to zero after $2n$ steps, $C_\tree(2n)=0$.
It should therefore not come as a surprise that with Brownian scaling we obtain the convergence in distribution (with respect to the uniform norm topology on real functions on the interval $[0,1]$) 
\begin{equation}\label{eq:contourconvergence}
    \left(t \mapsto \frac{C_\tree(2n t)}{c \sqrt{2n}}\right)\xrightarrow[n\to\infty]{(\mathrm{d})} \mathbf{e}
\end{equation}
where $\mathbf{e}:[0,1]\to \mathbb{R}_{\geq 0}$ is a \emph{Brownian excursion} \cite{Revuz1991} i.e.\ a standard Brownian motion started at $\mathbf{e}(0)=0$ and conditioned to stay non-negative until returning to zero after unit time, $\mathbf{e}(1)=0$.

\begin{figure}[t]
    \centering
    \includegraphics[width=\linewidth]{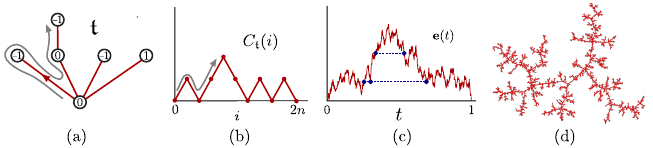}
    \caption{(a) The plane tree $\tree$ from Fig.~\ref{fig:cvsbijection}. (b) Its contour function $C_\tree(i)$. (c) The Brownian excursion $\mathbf{e}(t)$ and examples of a pair and a triple of points that is identified by $d_{\mathbf{e}}$. (d) An illustration of the CRT.}
    \label{fig:contour}
\end{figure}

Any continuous excursion $X : [0,1]\to \mathbb{R}_{\geq 0}$ naturally gives rise to a continuous metric space called a \emph{real tree}.
This is achieved by considering the metric\footnote{More precisely, this determines a \emph{pseudo-metric} on $[0,1]$. If we consider the equivalence relation $s \sim t$ when $d_X(s,t)=0$, then $d_X$ descends to a proper metric on the quotient $[0,1]/\sim$.}
\begin{equation}
    d_X(s,t) = X(s) + X(t) - 2 \inf_{u\in[s,t]} X(u), \qquad 0\leq s \leq t\leq 1
\end{equation}
on $[0,1]$, which identifies two points $s,t$ at the same height whenever $X$ does not drop below that height between $s$ and $t$ (see Fig.~\ref{fig:contour}c), i.e.\ $X(s)=X(t)$ and $X(u) \geq X(s)$ for all $u \in [s,t]$.
In the case of a Brownian excursion $X = \mathbf{e}$, this random metric space defines the Continuum Random Tree. 
Moreover, the convergence \eqref{eq:contourconvergence} of contour functions implies that the random tree $\mathfrak{t}$ converges in distribution, upon rescaling its graph distance by $1/\sqrt{2n}$, to the CRT in the Gromov--Hausdorff topology \cite{LeGall2005}.

The convergence \eqref{eq:contourconvergence} can be obtained much more generally for random plane trees, including the random mobiles associated to generic critical $\mathbf{q}$-Boltzmann maps \cite[Sec.~4]{Marckert2007}.
Indeed, one may interpret the CRT as a universal scaling limit of random tree-like geometries, which is known in the physics literature as the \emph{branched polymer universality class} \cite{DeGennes1979,Cates1985,Ambjorn_Diseases_1985,Ambjorn_1986}.

\subsection{Definition of the Brownian sphere}\label{sec:browniansphere}

Geodesic distances in the quadrangulations, at least towards the origin, are encoded in the labels of the trees.
Once the tree is known they satisfy a very simple law: the increments along the edges are independent and uniform in $\{-1,0,1\}$. 
In particular, if one examines the labels along a single path of vertices starting at the root and ending at a leaf of the tree, then the labels describe a random walk with steps in $\{-1,0,1\}$.
Since the length of a typical path is of order $\sqrt{n}$ the range of the random walk, and thus of the graph distances in the quadrangulation, is of order $\sqrt[4]{n}$, in accordance with the observations in Section~\ref{sec:distancestatistics}.
If we summarize the labels of the corners visited in the contour of $\tree$ by the label function $\ell_\tree : [0,2n] \to \mathbb{R}$, then one should thus expect the rescaled label function $t \mapsto \ell_\tree (2n t) / \sqrt[4]{n}$ to admit a limit as $n\to\infty$. 

This limit corresponds to the \emph{(head of the) Brownian snake} \cite{LeGall_Spatial_1999}, which informally amounts to a Brownian motion indexed by the branches of a CRT.
More precisely, given a Brownian excursion $\mathbf{e}$ that describes the contour of a CRT, we let $Z: [0,1]\to \mathbb{R}$ be the random continuous function with $Z(0)=Z(1)=0$, zero mean $\mathbb{E}[Z(t)]=0$, and Gaussian distribution determined by\footnote{Compare this with standard Brownian motion $B :\mathbb{R} \to \mathbb{R}$ on the line satisfying $\mathbb{E}[(B(s)-B(t))^2] = |s-t|$, which is the natural metric on $\mathbb{R}$.}
\begin{equation}
    \mathbb{E}[ (Z(s)-Z(t))^2] = d_{\mathbf{e}}(s,t).
\end{equation}

\begin{figure}[t]
    \centering
    \includegraphics[width=.7\linewidth]{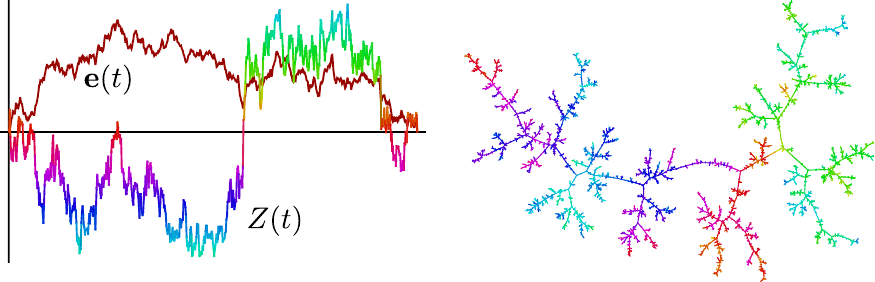}
    \caption{A simulation of the Brownian snake $(\mathbf{e},Z)$ illustrated as functions on $[0,1]$ (left) and as coloring of an embedding of $\tree$ in the plane (right). The colors in both figures match.}
    \label{fig:browniansnake}
\end{figure}

If $Z(t)$ attains its minimum at $t_0 = \operatorname{arg\,min}_{[0,1]} Z \in[0,1]$, then we wish to interpret $t_0$ as the origin (the analogue of the distinguished vertex in the map) and $Z(t) - Z(t_0)$ as the distance $D(t,t_0)$ from the $t$ to the origin. 
In order to define a metric space we should specify distances $D(s,t)$ between all pairs of points $s,t\in[0,1]$.
It is not immediately clear how to do this, but the inequality \eqref{eq:dgrinequality} derived back in Section~\ref{sec:treetomap}, suggests a natural upper bound on $D(s,t)$.
Namely, one should be able to follow the geodesics from $s$ and $t$ to the origin until they merge.
If $t_0 \notin [s,t]$ then the point of merging is $u=\operatorname{arg\,min}_{[s,t]} Z \in[0,1]$, and if $t_0 \in (s,t)$ then $u=\operatorname{arg\,min}_{[0,t]\cup[s,1]} Z \in[0,1]$.
The continuous analogue of \eqref{eq:dgrinequality} is then 
\begin{equation}
    D(s,t) \leq D^{\circ}(s,t) \coloneqq Z(s) + Z(t) -2 Z(u).
\end{equation}
However, $d_{\mathbf{e}}(t,t')=0$ does not imply $D^\circ(s,t)= D^\circ(s,t')$, essentially because some points will have multiple shortest geodesics to the origin, so $D^\circ(s,t)$ does not determine a metric on $[0,1]/\sim$. 
But it can be shown \cite{Marckert2006,LeGall_topological_2007} that there is a unique (largest) metric $D(s,t)$ satisfying the inequality, and that it is obtained by stringing together many pieces of geodesics to the origin,
\begin{equation}
    D(s,t) \coloneqq \inf \left\{ D^{\circ}(s,u_1) + D^{\circ}(v_1,u_2) + \cdots + D^{\circ}(v_k,t) : u_i \sim v_i\text{ for }1\leq i\leq k \right\}.
\end{equation} 
This random metric on $[0,1]$, with pairs of points $s,t$ identified whenever $D(s,t) = 0$, is called the Brownian map \cite{Marckert2006} or Brownian sphere $\mathbf{B} = ([0,1]/\sim,D)$.

\subsection{Properties of the Brownian sphere}\label{sec:Browniansphereproperties}

Even though it is a pretty wild metric space (judging by Fig.~\ref{fig:brownianmap}), the Brownian sphere is still a topological manifold.
Indeed, the metric $D(s,t)$ induces a topology on $[0,1]/\sim$ that was shown in \cite{LeGall_Scaling_2008} to be that of the $2$-sphere.
The anomalous scaling of geodesic distances with respect to areas, that we already observed in Section~\ref{sec:distancestatistics}, is reflected in the Brownian sphere having a Hausdorff dimension equal to $4$ almost surely \cite{LeGall_topological_2007}.

Besides the metric structure, the Brownian sphere possesses a natural volume measure coming from the Lebesgue measure on the interval $[0,1]$, which in particular provides a means of sampling uniform points in the surface. 
The root (corresponding to the endpoints of the interval $[0,1]$) and the origin (corresponding to $\operatorname{arg\,min}_{[0,1]}Z$) are such uniform points themselves.
The volume measure is precisely the scaling limit of the discrete measure that assigns equal volume to each vertex (or face) of the map, normalized to have total volume equal to one.
Proving this \cite{LeGall2019,Marzouk2022} requires a refinement of the convergence to the Gromov--Hausdorff--Prokhorov topology that also takes into account the structure provided by the measure.
Recently it has been demonstrated \cite{LeGall2022} that the volume measure does not provide extra information, but is completely determined by the metric.

\begin{figure}[t]
    \centering
    \includegraphics[width=.5\linewidth]{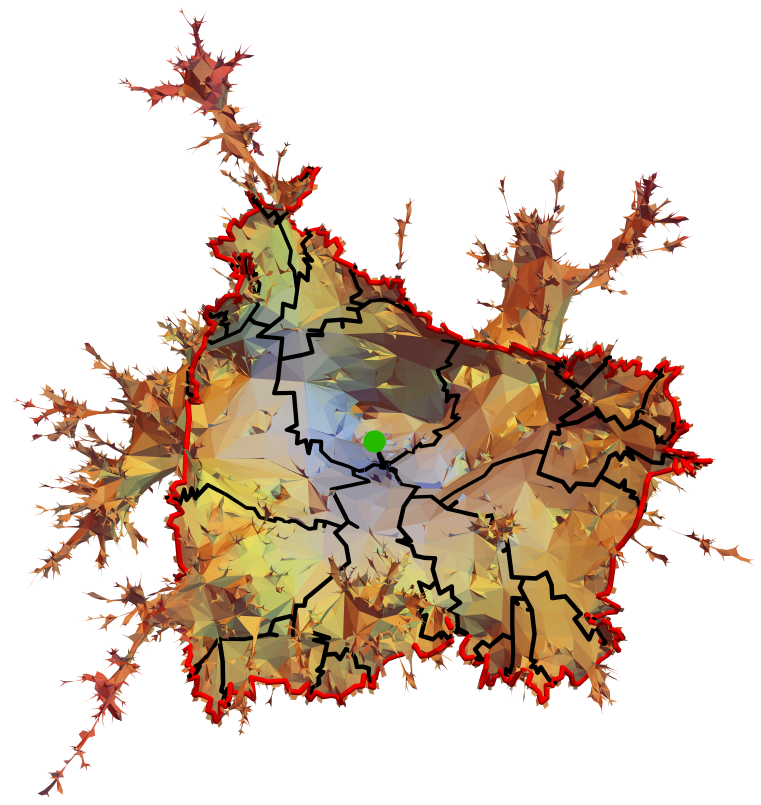}\includegraphics[width=.5\linewidth]{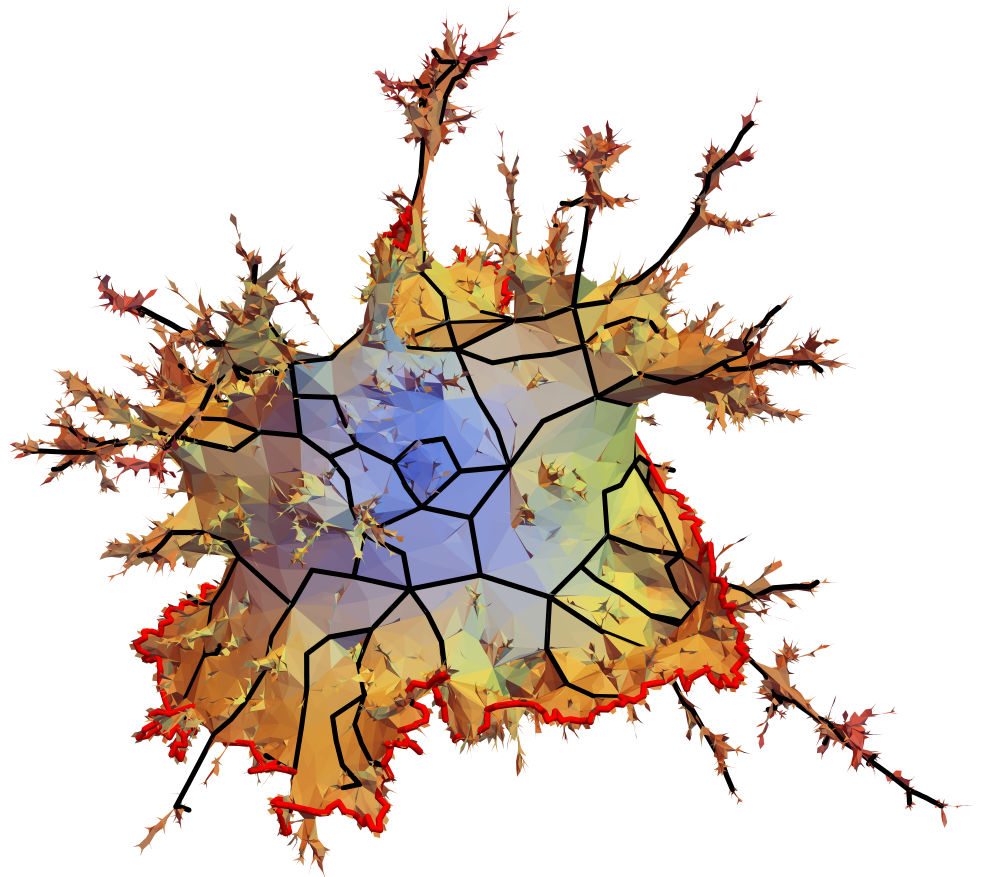}
    \caption{On the left a simulation of the hull of a geodesic ball of radius $r$ around the origin (green dot) in the Brownian sphere, where the boundary at distance $r$ is shown in red. A variety of geodesics (of length $r$) from boundary points to the origin is depicted in black, illustrating the merging of geodesics. On the right an illustration of the geodesic frame on the same surface, obtained by considering the union of geodesics between many pairs of (well-separated) random points.}
    \label{fig:browniangeodesics}
\end{figure}

To get a feeling for the fractal geometric of the Brownian sphere, it is instructive to examine its geodesics (see e.g.\ \cite{Bouttier2009,LeGall2010,Miermont_Brownian_2013,Angel2017,Miller2020,LeGall2022a}).
The Brownian sphere $\mathbf{B}$ is a geodesic space, in the sense that for any two points $x,y \in \mathbf{B}$ at distance $r=D(x,y)$ there is a continuous geodesic $\Gamma : [0,r] \to \mathbf{B}$ with $\Gamma(0)=x$ and $\Gamma(r)=y$ such that $D(\Gamma(s),\Gamma(t)) = t-s$ for $0\leq s\leq t\leq r$.
If $x$ and $y$ are sampled uniformly (from the volume measure), this geodesic is unique \cite{LeGall2010}.
This is similar to the situation in a Riemannian manifold, but the similarity stops when considering the structure of multiple geodesics.
We have already argued, based on the analysis of left-most geodesics in Section~\ref{sec:treetomap}, that the geodesics from $x$ and $y$ to a third uniform point $z$, say the origin, almost surely merge before reaching $z$ (Fig.~\ref{fig:browniangeodesics}).
This suggests that essentially there is just one way to approach a typical point $z$ via a geodesic route, contrary to Riemannian geometry where one can approach a point from any angle via a geodesic.
Of course, there will be atypical points $z\in\mathbf{B}$ that have more than one geodesic ending at $z$: for instance the interior points of geodesics have at least two, and the merger points of geodesics at least three.
In fact, it has been shown \cite{Miller2020,LeGall2022a} that the set of points of $\mathbf{B}$ where $m=1,2,3$ or $4$ geodesics meet has Hausdorff dimension $5-m$.
No such points exist where $m=6$ geodesics meet \cite{Miller2020}, while it is an open question whether there are points with $m=5$ geodesics \cite{Miller2020,LeGall2022a}. 
Perhaps even more strikingly, the \emph{geodesic frame} \cite{Angel2017} of $\mathbf{B}$, which is the union of all geodesics minus their endpoints between pairs of points in $\mathbf{B}$, has Hausdorff dimension one \cite{Miller2020}, equal to the dimension of a single geodesic.

\subsection{The Brownian sphere from Liouville Quantum Gravity}

The Brownian sphere, introduced to capture the scaling limit of random planar maps, gives a mathematically precise interpretation of what we would like to call pure quantum gravity on the 2-sphere. 
There is another way of constructing the same random metric space, starting from random Riemannian geometry on the 2-sphere, going under the name of Liouville Quantum Gravity, which has seen a flurry of activity in recent years in probability theory.
Summarizing these developments, and their connections to Liouville Conformal Field Theory and the Mating of Trees approach, goes far beyond the scope of this chapter, so we direct the reader to several key papers \cite{Duplantier2011,Sheffield2016,Miller2020a,Miller2021,Miller2021a,Duplantier2021,David2016,Gwynne2019,Ding2019,Ding2020,Gwynne2021} and review articles \cite{Gwynne2019a,Ding2021,Sheffield2022}.
Here we restrict to superficially describing the process to arrive at a metric space that is known to agree with the Brownian sphere.

The general idea, going back to Polyakov \cite{Polyakov1981} in the context of non-critical bosonic string theory, is to uniformize Riemannian metrics $g_{ab}$ on the $2$-sphere as conformal rescalings of a fixed background metric $\hat{g}_{ab}$,
\begin{equation}\label{eq:uniformization}
    g_{ab}(x) = e^{\gamma \phi(x)} \hat{g}_{ab}(x),
\end{equation}
and study the conformal field theory of the Liouville field $\phi$ for gravity, possibly coupled to conformal matter fields.
Liouville Quantum Gravity with Liouville coupling $\gamma\in(0,2)$ corresponds to the rigorous path integral quantization of the Liouville field $\phi$ with action
\begin{equation}\label{eq:liouvilleaction}
    S_{\mathrm{L}}[\phi] = \frac{1}{4\pi}\int_{S^2}\rmd^2 x\sqrt{\hat{g}}\left(\hat{g}^{ab}\partial_a\phi\partial_b\phi +Q\hat{R}\phi+4\pi\hat{\mu}e^{\gamma \phi}\right),
\end{equation}
where $\hat{\mu}>0$ is a cosmological constant, $\hat{R}$ the scalar curvature of $\hat{g}_{ab}$ and
\begin{equation}
    Q = \frac{2}{\gamma}+\frac{\gamma}{2}.
\end{equation} 
The value of $\gamma$ is determined by the central charge $c\in (-\infty,1)$ of the coupled matter system and is related to $Q\in(2,\infty)$ via
\begin{equation}
    c = 25 - 6 Q^2.
\end{equation}
Pure gravity thus corresponds to $c=0$, $Q = 5/\sqrt{6}$ and $\gamma=\sqrt{8/3}$.
For a rigorous construction of the path integral as a measure on an appropriate function space of fields $\phi: S^2 \to \mathbb{R}$ using so-called Gaussian Multiplicative Chaos, we refer the reader to \cite{Kahane_Sur_1985,Rhodes2014,David2016,Berestycki_elementary_2017}.
This is achieved by considering the measure as a deformation of the \emph{Gaussian Free Field} (GFF) on the 2-sphere with metric $\hat{g}_{ab}$, which is the random massless scalar field with action corresponding to the quadratic part of \eqref{eq:liouvilleaction}.

In connection with the (unit-volume) Brownian sphere, we are interested in the unit-volume Liouville field, which can be shown \cite{David2016,Aru2017} to be related to the GFF by a manageable deterministic position-dependent shift.
For simplicity, we consider the Riemann sphere $\mathbb{C}\cup\{\infty\} \cong S^2$ with the flat Euclidean background metric $\hat{g}_{ab} = \delta_{ab}$ on $\mathbb{C}\cong \mathbb{R}^2$.
Since the GFF and $\phi$ are logarithmically correlated, $\mathbb{E}[\phi(x)\phi(y)] \sim \log 1/|x-y|$ as $x\to y$, the random field cannot be point-wise defined and regularization is necessary to make sense of the exponential $e^{\gamma \phi(x)}$.
This can be achieved by averaging $\phi$ over a neighborhood of small radius $\varepsilon>0$, for instance by taking $\phi_\varepsilon(x)$ to be a version of $\phi$ mollified by the heat kernel,
\begin{equation}
    \phi_\varepsilon(x) = \int_\mathbb{C} \rmd^2 y  \frac{1}{\pi \varepsilon^2}e^{- \frac{|x-y|^2}{\varepsilon^2}}\phi(y).
\end{equation}
Then $\phi_\varepsilon(x)$ is a nice continuous random function that one can exponentiate to give a random Riemannian metric 
\begin{equation}
    g^\varepsilon_{ab} = e^{\gamma \bar{\phi}_\varepsilon(x)} \delta_{ab}, \qquad \bar{\phi}_\varepsilon(x) \coloneqq \phi_\varepsilon(x) - \frac{1}{\gamma}\log\left(\int_{S^2} e^{\gamma \phi_\varepsilon(x)}\,\rmd^2 x\right).
\end{equation}
of unit volume, $\int_{S^2} \sqrt{g^\varepsilon}\rmd^2 x = 1$.
The unit-volume \emph{Liouville quantum measure} $\mu_\phi$ on $S^2 \cong \mathbb{C} \cup \{\infty\}$ is then defined as the limiting measure \cite{Kahane_Sur_1985,Duplantier2011,Rhodes2014,Berestycki_elementary_2017},
\begin{equation}
    \mu_\phi = \lim_{\varepsilon\to 0} e^{\gamma \bar{\phi}_\varepsilon(x)}\,\rmd^2 x.
\end{equation}
It is independent of the chosen background metric $\hat{g}_{ab}$ and transforms covariantly under conformal transformations of $S^2$ \cite{Duplantier2011}.
In the pure-gravity case $\gamma=\sqrt{8/3}$ this random measure should correspond to the measure on the Brownian sphere, that we discussed in Section~\ref{sec:Browniansphereproperties}.

The usual metric space structure one would associate to a Riemannian metric like $g_{ab}^\varepsilon$ would be based on the shortest length $\int_0^1 \rmd t\sqrt{g_{ab}^\varepsilon\dot{\Gamma}^a\dot{\Gamma}^b} =\int_0^1 \rmd t|\dot{\Gamma}|e^{\frac{\gamma}{2} \bar{\phi}_\varepsilon(\Gamma(t))}$ of paths $\Gamma:[0,1]\to S^2$ between two points, resulting in the geodesic distance
\begin{equation}
    d_{g^{\varepsilon}}(x,y) = \inf\left\{\int_0^1 \rmd t|\dot{\Gamma}|e^{\frac{\gamma}{2} \bar{\phi}_\varepsilon(\Gamma(t))}:\text{paths }\Gamma\text{ from }\Gamma(0)=x\text{ to }\Gamma(1)=y \right\}.
\end{equation}
However, this \emph{cannot} be the right answer!
Shifting the field $\phi$ in a small neighborhood by a constant $c$ leads to a local increase of volume, as measured by $\mu_\phi$, by a factor $e^{\gamma c}$, while local geodesic distances $d_{g^\varepsilon}(x,y)$ scale by a factor $\sqrt{e^{\gamma c}}$. 
This is at odds with the anomalous scaling of these quantities in the Brownian sphere, in which they should differ by a power of $4$, i.e.\ the Hausdorff dimension of the metric space.

More generally, each value of the coupling constant $\gamma \in (0,2)$ is expected to correspond to a universality class of two-dimensional quantum gravity coupled to conformal matter and have an associated Hausdorff dimension $d_\gamma > 2$ with $d_{\sqrt{8/3}}=4$.
So the appropriate definition, going under the name of \emph{Liouville first passage percolation} \cite{Ding_Liouville_2019,Dubedat_Weak_2020,Ding2020,Gwynne2021}, is to set
\begin{equation}
    \xi = \frac{\gamma}{d_{\gamma}}    
\end{equation}
and consider instead the metric
\begin{equation}
    D^\varepsilon_{\phi}(x,y) = \inf\left\{\int_0^1 \rmd t|\dot{\Gamma}|e^{\xi\bar{\phi}_\varepsilon(\Gamma(t))}:\text{paths }\Gamma\text{ from }\Gamma(0)=x\text{ to }\Gamma(1)=y \right\}.
\end{equation}
It has been demonstrated \cite{Ding2020,Dubedat_Weak_2020,Gwynne2021} that $D_\phi^\varepsilon(\cdot,\cdot) / a_\varepsilon$ with an appropriate normalization $a_\varepsilon \approx \varepsilon^{1-\xi Q}$ converges as $\varepsilon\to 0$ (in probability with respect to an appropriate topology) to a random metric space on $S^2$.
Moreover, in the case $\gamma=\sqrt{8/3}$ it coincides \cite{Gwynne2021} up to a global rescaling with the metric constructed from $\phi$ via \emph{Quantum Loewner Evolution} \cite{Miller2020a,Miller2021,Miller2021a}, which in turn has the same law as the Brownian sphere from Section~\ref{sec:browniansphere} \cite{Miller_axiomatic_2021,Miller2021}.
An even more precise link between the Brownian sphere and Liouville Quantum Gravity has been obtained in \cite{Holden2019} by establishing scaling limits of both the measure and the metric space with respect to a certain discrete conformal embedding of uniform random triangulations in the Euclidean plane, which forms a discrete counterpart of the uniformization \eqref{eq:uniformization}.

\section{Local limits}\label{sec:locallimits}

Before moving on to the complementary method of peeling explorations, this is a good moment to introduce a framework in which we can deal with infinite random maps via local limits.
In a sense this is a problem analogous to that of defining statistical systems, like the Ising model, on a fixed infinite lattice, where Gibbs measures \cite{Dobruschin1968,Lanford1969} play an important role. 
Just like Ising configurations on an infinite lattice, there are uncountably many infinite maps, so we need to start by introducing a convenient topology. 

\begin{figure}[t]
    \centering
    \sidecaption[t]
    \includegraphics[width=.58\linewidth]{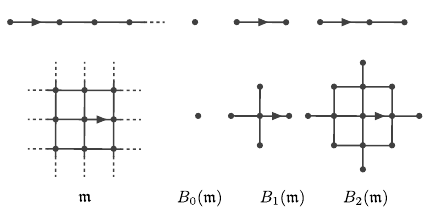}
    \caption{Two examples of one-ended infinite planar maps $\map$ together with their balls $B_r(\map)$ of increasing radius with respect to the graph distance.}
    \label{fig:infmaps}
\end{figure}

Following the foundational work of Benjamini and Schramm \cite{Benjamini2001}, let us introduce the \emph{local topology}, in which, informally, two rooted maps are close to each other if they are identical in a large neighbourhood of the root.
More precisely, if $\map$ is a rooted map, we let the \emph{ball $B_r(\map)$ of radius $r$} be the subset of $\map$ consisting of all vertices at graph distance at most $r$ from the start of the root edge and all edges that have at least one of their endpoints at distance smaller than $r$.
Then the \emph{local distance} between two maps $\map$ and $\map'$ is defined to be 
\begin{equation}
    d_{\mathrm{Loc}}(\map,\map') = \frac{1}{1+ \sup \{r\geq 0 : B_r(\map) = B_r(\map')\}}.
\end{equation}
Any other strictly decreasing function approaching zero instead of the reciprocal could have been used, the point being that $d_{\mathrm{Loc}}$ satisfies the triangle inequality and $d_{\mathrm{Loc}}(\map,\map')=0$ if and only if $\map=\map'$, and therefore defines a metric on the space of all finite rooted planar maps $\mathcal{M}$.
It is not a complete metric (some Cauchy sequences do not have a limit), but becomes one when we add infinite rooted planar maps $\overline{\mathcal{M}} = \mathcal{M} \cup \mathcal{M}_\infty$.
Here we could define an \emph{infinite map} $\map\in \mathcal{M}_\infty$ as a sequence of finite maps $\map_0,\map_1,\map_2,\ldots \in \mathcal{M}$ representing the balls of increasing radius in $\map$, meaning that $B_r(\map_j) = \map_r$ for all $j\geq r\geq 0$. 
With this definition, all vertices of $\map$ must have finite degree, but $\map$ can have faces of infinite degree.
The simplest example of this is the map which has an infinite sequence of edges heading away from the root edge (top of Fig.~\ref{fig:infmaps}), which has a single face of infinite degree. 
It is an example of a \emph{one-ended} infinite planar map, which is a planar map for which the removal of any finite subset of edges results in connected components of which exactly one is infinite. 
Another example of a one-ended infinite planar map is the square grid with vertex set $\mathbb{Z}^2 \subset \mathbb{R}^2$ (bottom of Fig.~\ref{fig:infmaps}).
Just like a (finite) planar map can be viewed as a graph that is properly embedded in the sphere, a one-ended infinite planar map is an infinite graph that is properly embedded in the plane, in a locally finite fashion with its end at infinity \cite[Sec.~2.1]{Curien2019}.

The local distance induces a topology on $\overline{\mathcal{M}}$ that provides a notion of \emph{local convergence} (in distribution) of random planar maps to a random infinite map.
Suppose $\map^{(1)}, \map^{(2)}, \ldots$ are random (finite or infinite) planar maps and $\map$ is a random infinite planar map, then local convergence of $\map^{(n)}$ to $\map$ as $n \to \infty$ is equivalent to convergence in distribution of the balls of all radii, i.e.\ $B_r(\map^{(n)}) \xrightarrow[n\to\infty]{\mathrm{(d)}} B_r(\map)$ for all $r \geq 0$.

\begin{figure}[t]
    \centering
    \includegraphics[width=.65\linewidth]{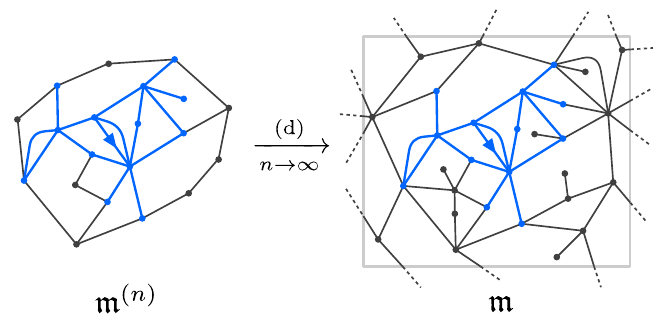}
    \caption{Illustration of the local convergence of a $\mathbf{q}$-Boltzmann planar map $\mathfrak{m}^{(n)}$ with $n$ edges to the $\mathbf{q}$-IBPM $\map$. For this it is necessary that for any fixed rooted map $\mathfrak{b}$, like the one indicated in thick blue, the probability of $\mathfrak{b}$ occurring as the ball of radius $r$ converges as $n\to\infty$, i.e.\ $\lim_{n\to\infty}\mathbb{P}(B_r(\map^{(n)}) = \mathfrak{b})=\mathbb{P}(B_r(\map) = \mathfrak{b})$.}
    \label{fig:ibpm}
\end{figure}

Many models of random planar maps admit such a local convergence.
The first result in this direction was the local convergence as $n\to\infty$ of the uniform planar triangulation with $2n$ triangles to the \emph{uniform infinite planar triangulation (UIPT)} obtained by Angel and Schramm in \cite{Angel_Uniform_2003}.
Similarly, uniform quadrangulations were shown to converge to the \emph{uniform infinite planar quadrangulation (UIPQ)} \cite{Krikun_Local_2006,Chassaing_Local_2006,
Menard_two_2010,Curien_view_2013}.
This was extended to the case of bipartite Boltzmann maps conditioned on the number of edge $n$ by Bj\"ornberg and Stef\'ansson in \cite{Bjoernberg_Recurrence_2014} (and later to general Boltzmann maps by Stephenson in \cite{Stephenson_Local_2018}).
To be precise, if $\mathbf{q}$ is a critical weight sequence and $\map^{(n)}$ is a (rooted) $\mathbf{q}$-Boltzmann planar map conditioned to have $n$ edges, then $\map^{(n)}$ converges locally to a unique one-ended random infinite map $\map$ called the \emph{infinite Boltzmann planar map ($\mathbf{q}$-IBPM)} (Fig.~\ref{fig:ibpm}).

\begin{figure}[t]
    \centering
    \includegraphics[width=.9\linewidth]{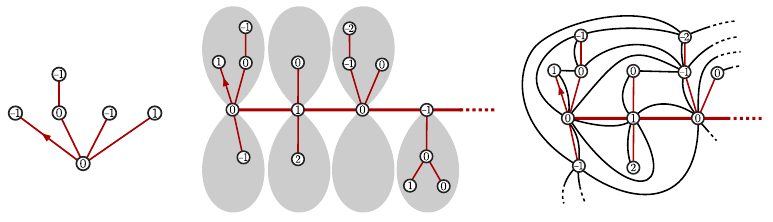}
    \caption{The local limit of a uniform (labeled) plane tree with $n$ edges (left) is an infinite tree consisting of a line with independent critical plane trees (shaded in gray) growing from both sides (middle). Applying the rules of the inverse Cori--Vauquelin--Schaeffer bijection to the infinite tree produces the UIPQ.}
    \label{fig:inftree}
\end{figure}

This local convergence can be understood from the point of view of the tree bijections of the previous section.
Let us illustrate this in the case of the UIPQ following \cite{Curien_view_2013}.
Recall the bijective encoding of rooted, pointed quadrangulations $\map^{(n)}$ with $n$ faces by labeled plane trees $\tree^{(n)}$ with $n$ edges of Section~\ref{sec:distancestatistics} (and Fig.~\ref{fig:cvsbijection}).
One can imagine that a local neighborhood of the root in the quadrangulation $\map^{(n)}$ is typically determined by a local neighborhood in the corresponding tree $\tree^{(n)}$, suggesting that one should consider the local limit of the random labeled tree first.
The latter, when forgetting the labels for a moment is an example of a Bienayme-Galton-Watson tree conditioned on its size, for which general local limits have been established by Kesten \cite{Kesten_Subdiffusive_1986}.
The limit corresponds to a one-ended infinite tree $\mathfrak{t}$ consisting of a \emph{spine}, i.e.\ an infinite line of vertices starting at the root vertex, with independent critical plane trees growing out on both sides (Fig.~\ref{fig:inftree}).
The labels have independent increments along the edges that are uniform in $\{-1,0,1\}$ and such that the root vertex has label $0$.
Since the labels along the spine describe a random walk with no drift, the range of the labels is almost surely the whole of $\mathbb{Z}$.
Applying the rules described in Section~\ref{sec:treetomap} to $\mathfrak{t}$, where we skip the first step (since there is no minimal label, so we do not need to add a new vertex to become the origin), results in an infinite quadrangulation (Fig.~\ref{fig:inftree}).
The result is one-ended and can be shown to be the local limit of $\map^{(n)}$, and therefore to describe the UIPQ \cite{Curien_view_2013}.
Note that the origin of $\map^{(n)}$, the distinguished vertex used to construct the distance labeling, does not appear in the limiting UIPQ anymore: in a sense it has drifted away to infinity in the limit.

As we will see in more detail in the next section, when discussing the peeling exploration, infinite random maps are particularly useful when discussing scaling properties.
For instance, in the UIPQ the expected number of vertices at distance $r$ grows like $r^3$ as $r\to\infty$ \cite{Chassaing_Local_2006}, nicely reflecting the Hausdorff dimension of $4$.

Starting from random infinite maps one can again consider continuum limits, the difference with before being that one does not have to worry about the size of the map and the scaling only has to be applied to the distances.
A convergence of this type has been established for the UIPQ with respect to a local version of the Gromov--Hausdorff topology in \cite{Curien2014}, and the limit is called the \emph{Brownian plane} because it has the topology of $\mathbb{R}^2$.
The Brownian plane is an example of a random metric space with exact scaling symmetry, in the sense that its distribution is unchanged when all distances are multiplied by a positive constant.
It can also be obtained by considering the infinite-volume limit of the Brownian sphere \cite{Curien2014}.

\section{The peeling process}\label{sec:peeling}

\begin{figure}[t]
    \centering
    \includegraphics[width=.75\linewidth]{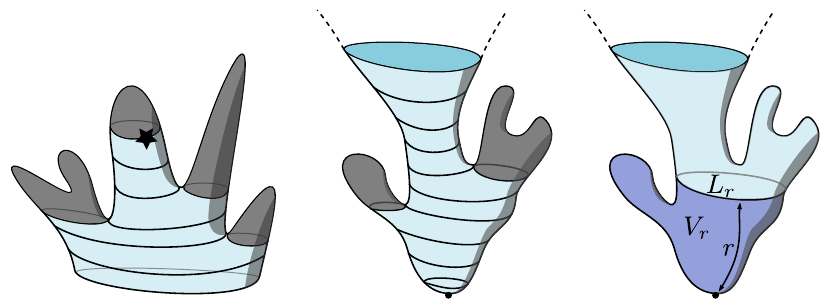}
    \caption{An intuitive picture of the peeling exploration of a pointed map starting at the boundary (left) and of an infinite map starting at the root (middle). We will see how to use the exploration to track the boundary length $L_r$ and area $V_r$ of geodesic ball of radius $r$ around the root (right).}
    \label{fig:peelingillustration}
\end{figure}

We have seen that bijections with labeled trees provide an explanation for the universal properties of planar map enumeration, while providing an economical way to study statistics of geodesic distances. 
In this section we will discuss a complementary approach that displays the universality in a different way and gives access to other classes of statistics.
This approach goes under the umbrella name of \emph{peeling}, which amounts to analyzing an exploration process on a random map (see Fig.~\ref{fig:peelingillustration} for an illustration).
Such a peeling process was first described by Watabiki \cite{Watabiki_Construction_1995} in the setting of the Euclidean Dynamical Triangulation approach to non-critical string theory.
It formed the basis for the first calculation of the geodesic two-point function \eqref{eq:geodtwopoint} by Ambj\o rn and Watabiki \cite{Ambjorn_Scaling_1995}.
The first appearance of peeling in the mathematics literature was in the work of Angel \cite{Angel_Growth_2003} on percolation on the uniform infinite planar triangulation (see Section~\ref{sec:locallimits}), which sparked many related investigations (for example \cite{Benjamini_Simple_2013,Angel_Percolations_2014,Richier_Universal_2015,Ambjorn_Multi_2016}). 
We will focus on a type of peeling process, going under the name of \emph{lazy peeling} or \emph{edge peeling}, that is particularly convenient for the study of Boltzmann planar maps.
It was formulated in \cite{Budd_Peeling_2016} and has been used as a tool to investigate many types of statistics related to these maps (for example \cite{Bertoin_Martingales_2017,Budd_Geometry_2017,Curien_Infinite_2021,Budzinski_Local_2022}).
For an in depth discussion and many applications we direct the reader to the lecture notes on the topic by Curien \cite{Curien2019}.

\subsection{Peeling explorations}

\begin{figure}[t]
    \centering
    \sidecaption[t]
    \includegraphics[width=.63\linewidth]{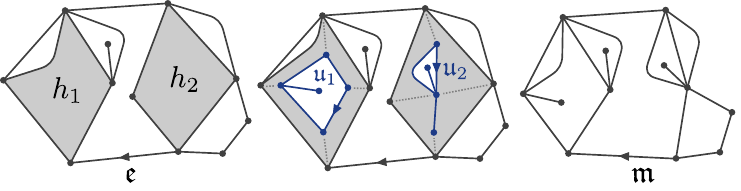}
    \caption{The planar map $\mathfrak{e}$ with two holes $h_1$ and $h_2$ is a submap of $\map$, because gluing $\mathfrak{u}_1$ and $\mathfrak{u}_2$ into the holes yields $\map$.}
    \label{fig:holes}
\end{figure}

Let $\map$ be a (rooted) bipartite planar map. 
The intuitive idea of (lazy) peeling of $\map$ is that we explore $\map$ one edge at a time starting from the root face until we have seen the entire map.
To formalize this we require a way to encode what part of $\map$ has been explored at each step.
To this end we introduce a \emph{planar map with holes} to be a planar map $\mathfrak{e}$ (the ``$\mathfrak{e}$xplored'' part) with a distinguished set of faces not including the root face, that we call the \emph{holes} of $\mathfrak{e}$ (left side of Fig.~\ref{fig:holes}).
Each hole is required to be \emph{simple}, meaning that its contour does not visit any vertex twice, and the holes are not allowed to touch each other.
Given a hole $h$ of degree $2k$ and a planar map $\mathfrak{u}$ (the ``$\mathfrak{u}$nexplored'' part) with root face of degree $2k$, we have a natural operation\footnote{Note that to make this operation unambiguous one should fix an algorithm to select an edge in the contour of hole $h$ to which the root edge of $\mathfrak{u}$ is to be glued. There are many choices for such an algorithm, but since it will not affect any of the further considerations, we will leave it unspecified.} of \emph{gluing} $\mathfrak{u}$ into the hole $h$ of $\mathfrak{e}$, which is best explained in a picture, see Fig.~\ref{fig:holes}.

If $\mathfrak{e}$ is a planar map with $p\geq 0$ holes $h_1,\ldots,h_p$, then $\mathfrak{e}$ is said to be a \emph{submap} of $\map$, denoted $\mathfrak{e} \subset \map$, if planar maps $\mathfrak{u}_1,\ldots,\mathfrak{u}_p$ exist such that the result of gluing $\mathfrak{u}_i$ into the hole $h_i$ is $\map$.
More generally, if $\mathfrak{e}'$ is another planar map with holes, we can make sense of $\mathfrak{e}$ being a submap of $\mathfrak{e}'$, by allowing the maps $\mathfrak{u}_i$ to have holes themselves.
Importantly, we may convince ourselves that as soon as $\mathfrak{e}\subset \mathfrak{e}'$, the maps $\mathfrak{u}_i$ that need to be glued in the holes of $\mathfrak{e}$ are uniquely determined.
We will call the edges and vertices of $\mathfrak{e}$ that are adjacent to a hole \emph{active} and the other ones \emph{explored}.

\begin{figure}[t]
    \centering
    \sidecaption[t]
    \includegraphics[width=.6\linewidth]{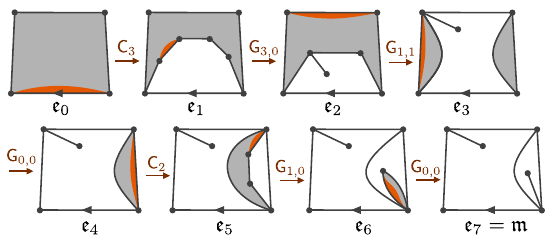}
    \caption{Example of a peeling exploration of a map $\map$ with $7$ edges. The peel edge selected by the (unspecified) algorithm $\mathcal{A}$ is shown with orange shading, and the peeling transitions $\mathsf{C}_k$ or $\mathsf{G}_{k_1,k_2}$ are indicated.}
    \label{fig:untargetedpeeling}
\end{figure}

A \emph{lazy peeling} or \emph{edge peeling} exploration of $\map$ then is an increasing sequence of submaps
\begin{equation}\label{eq:exploration}
    \mathfrak{e}_0 \subset \mathfrak{e}_1 \subset \mathfrak{e}_2 \subset \cdots \subset \mathfrak{e}_{|\edges(\map)|} = \map,
\end{equation}
where $\mathfrak{e}_i$ has precisely $i$ internal edges.
In particular, the initial submap $\mathfrak{e}_0\subset \map$ has just two faces, the root face and a hole of the same degree. See Fig.~\ref{fig:untargetedpeeling} for an example.
This definition is a bit abstract, but should become more clear if we analyze the possible transitions $\mathfrak{e}_i \subset \mathfrak{e}_{i+1}$. 
This transition is a result of the operation of \emph{peeling an edge} $e$ where $e$ is an active edge of $\mathfrak{e}_i$ (indicated by orange shading in Fig.~\ref{fig:untargetedpeeling}).
Note that $e$ corresponds to a unique (side of an) edge in $\map$, and let $f$ be the face of $\map$ that sits on the other side of this edge.
Then we distinguish two types of events depending on whether $f$ is new to $\mathfrak{e}_i$ or not:
\begin{itemize}
    \item Event $\mathsf{C}_k$, when $f$ was not already in $\mathfrak{e}_i$ and its degree is $2k$. Then $\mathfrak{e}_{i+1}$ is obtained from $\mathfrak{e}_i$ by attaching a $2k$-gon to $e$ inside the hole.
    \item Event $\mathsf{G}_{k_1,k_2}$, when $f$ was already present in $\mathfrak{e}_i$. In this case $\mathfrak{e}_{i+1}$ is obtained from $\mathfrak{e}_i$ by gluing $e$ to another edge $e'$ in the contour of the same hole, splitting the hole into two holes of degrees $2k_1\geq 0$ and $2k_2\geq 0$ (the first on the right of $e$, the second on the left). We can have $k_1 =0$ or $k_2=0$ when $e$ and $e'$ are adjacent, in which case the corresponding hole is not really a hole but a single vertex.
\end{itemize}
In particular, given $\mathfrak{e}_i \subset \map$ the edge $e$ uniquely determines the result $\mathfrak{e}_{i+1}$.
So if we choose a \emph{peeling algorithm} $\mathcal{A}$ that chooses from any planar map with holes an active edge $e = \mathcal{A}(\mathfrak{e})$ to peel, then $\mathcal{A}$ and $\map$ together uniquely specify the exploration \eqref{eq:exploration}.
The versatility of the peeling approach lies in the freedom one has in specifying the algorithm $\mathcal{A}$, while the results that follow are independent of this choice.

\subsection{Targeted peeling of a pointed or infinite planar map}

It is often useful to consider a targeted version of the peeling exploration.
The \emph{target}, depicted by a star $\star$, can be the distinguished vertex in a pointed map or the boundary at infinity in a one-ended infinite map.
In the former case the exploration process stops when the target is explored, while evidently in the later case the exploration continues indefinitely.
Either way, it makes sense to speed up the exploration process, by filling in a newly produced hole $h$ with its corresponding unexplored region $\mathfrak{u}$ of $\map$ whenever $\mathfrak{u}$ does not contain the target. 
In this way the peeling exploration becomes a sequence of submaps 
\begin{equation}
    \bar{\mathfrak{e}}_0 \subset \bar{\mathfrak{e}}_1 \subset \cdots \subset \map,
\end{equation}
where $\bar{\mathfrak{e}}_0 = \mathfrak{e}_0$ as before and each of the submaps $\bar{\mathfrak{e}}_0,\ldots, \bar{\mathfrak{e}}_{n-1}$ has a single hole.
The transitions can be deduced from those of the untargeted peeling.
In the event $\mathsf{C}_k$ no hole needs to be filled in.
In case of $\mathsf{G}_{k_1,k_2}$, there are two possibilities: either the hole of degree $2k_1$ is filled in, an event that we denote by $\mathsf{G}_{k_1,\star}$, or the hole of degree $2k_2$ is filled in, denoted by $\mathsf{G}_{\star,k_2}$.
Note that the final step $\bar{\mathfrak{e}}_{n-1} \subset \bar{\mathfrak{e}}_{n} = \map$ in the pointed case necessarily corresponds to an event $\mathsf{G}_{\star,p-1}$ or $\mathsf{G}_{p-1,\star}$ where $2p$ is the degree of the hole of $\bar{\mathfrak{e}}_{n-1}$, because only when the two active edges adjacent to the target are glued, the target is explored.

\begin{figure}[t]
    \centering
    \includegraphics[width=\linewidth]{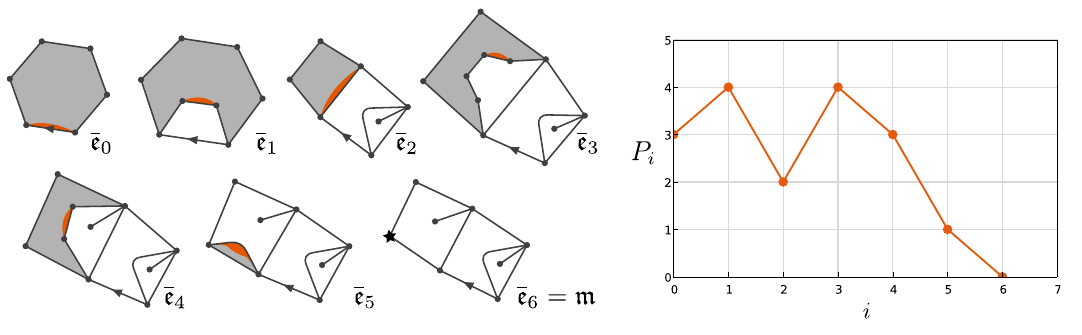}
    \caption{Example of a targeted peeling exploration of a pointed map $\map$ together with the corresponding perimeter process $(P_i)_{i\geq0}$. The peel edge selected by the (unspecified) algorithm $\mathcal{A}$ is indicated with the orange shading.}
    \label{fig:targetedpeeling}
\end{figure}

To a targeted peeling exploration one may naturally associate the sequence $(P_i)_{i\geq 0}$ of integers, called the \emph{perimeter process}, by setting $P_i$ to be half the degree of the hole of $\mathfrak{e}_i$.
Then $P_0$ is equal to half the perimeter of $\map$, and $P_i>0$ for all $i > 0$ in case $\map$ is an infinite map, while $P_i > 0$ for $i < n$ and $P_n=0$ for a pointed map $\map = \bar{\mathfrak{e}}_n$.
The increments $P_{i+1}-P_i$ are determined by the peeling event leading from $\bar{\mathfrak{e}}_i$ to $\bar{\mathfrak{e}}_{i+1}$.
Indeed, we see that 
\begin{equation}\label{eq:perimeterincrements}
    P_{i+1} - P_i = \begin{cases}
        k-1 &\text{in the event }\mathsf{C}_k,\\
        -k-1 & \text{in the event }\mathsf{G}_{\star,k}\text{ or }{G}_{k,\star}.
    \end{cases}
\end{equation}

\subsection{Peeling pointed Boltzmann planar maps}

Now suppose $\mathbf{q}$ is admissible and $\map$ is a pointed $\mathbf{q}$-Boltzmann planar map with specified root face degree $2\ell$. 
In other words, we consider the probability distribution 
\begin{equation}\label{eq:pointeddistrib}
    \mathbb{P}^{(2\ell)}_{\bullet}(\map) \coloneqq \frac{w_\mathbf{q}(\map)}{W_\bullet^{(2\ell)}},
\end{equation}
with the weight $w_\mathbf{q}(\map)$ as in \eqref{eq:diskdefinition}.
If we fix a peeling algorithm $\mathcal{A}$, then $\map$ determines a random peeling exploration $\bar{\mathfrak{e}}_0 \subset \bar{\mathfrak{e}}_1 \subset \cdots \subset \map$, which has a simple description.
The reason for the simplicity is the following \emph{domain Markov property} following from the factorized form the distribution \eqref{eq:pointeddistrib}: for any $i$, conditionally on $\mathfrak{e}_i$, the unexplored region $\mathfrak{u}$ corresponding to the hole of degree $2p$ of $\mathfrak{e}_i$ has distribution $\mathbb{P}^{(2p)}_{\bullet}$. 
From this we deduce that the event $\mathsf{C}_k$ occurs with probability
\begin{equation}
    \frac{q_{2k}W_\bullet^{(2p+2k-2)}}{W_\bullet^{(2p)}}
\end{equation}
and the events $\mathsf{G}_{\star,k}$ and $\mathsf{G}_{k,\star}$ each with probability
\begin{equation}
    \frac{W_\bullet^{(2p-2k-2)}W^{(2k)}}{W_\bullet^{(2p)}}.
\end{equation}
That these probabilities add up to one, can be checked from taking a $t$-derivative of the Tutte equation \eqref{eq:tutte} and using $W_\bullet^{(2\ell)} = \frac{\partial}{\partial t}W^{(2\ell)}$ before setting $t=1$.

It follows from \eqref{eq:perimeterincrements} that the perimeter process $(P_i)_{i=0}^n$ becomes a Markov process on the non-negative integers with transition probabilities
\begin{align}
    \mathbb{P}_\bullet^{(2\ell)}( P_{i+1} = p+k | P_i = p) &= \frac{W_\bullet^{(2p+2k)}}{W_\bullet^{(2p)}} \begin{cases}
        q_{2k+2} & \text{if }k \geq 0 \\ 2 W^{(-2k-2)} & \text{if }-p \leq k \leq -1,
    \end{cases}
\end{align}
where the process stops at the first time $i$ for which $P_i=0$, i.e.\ when $\bar{\mathfrak{e}}_i = \map$. 
Recalling the universal form \eqref{eq:pointedevendisk} of the pointed disk function and introducing the notation
\begin{equation}\label{eq:hdownandnu}
    h^\downarrow(\ell) = 4^{-\ell} \binom{2\ell}{\ell} \ind_{\{\ell \geq 0\}}, \qquad \nu_{\mathbf{q}}(k) = \begin{cases}
        q_{2k+2} (4R)^k& \text{if }k \geq 0 \\ 2 W^{(-2k-2)} (4R)^k & \text{if } k \leq -1,
    \end{cases}
\end{equation}
the transition probabilities can be summarized as
\begin{align}\label{eq:perimeterlaw}
    \mathbb{P}_\bullet^{(2\ell)}( P_{i+1} = p+k | P_i = p) &= \frac{h^\downarrow(p+k)}{h^\downarrow(p)} \nu_{\mathbf{q}}(k).
\end{align}
Using \eqref{eq:diskandpotential} and \eqref{eq:onecutdisksolution}, we may check that
\begin{equation}
    \sum_{k=-\infty}^\infty \nu_\mathbf{q}(k) \stackrel{\eqref{eq:diskandpotential}}{=} 1 - \frac{V'(\sqrt{4R})}{\sqrt{4R}} + 2 \frac{W(\sqrt{4R})}{\sqrt{4R}} \stackrel{\eqref{eq:onecutdisksolution}}{=} 1.
\end{equation}
Therefore, $\nu_\mathbf{q}$ determines a probability measure on $\mathbb{Z}$, corresponding to the distribution of increments of the perimeter process in the large-perimeter limit, i.e.\ the large-$p$ limit of \eqref{eq:perimeterlaw}.

\begin{figure}[t]
    \centering
    \sidecaption[t]
    \includegraphics[width=.5\linewidth]{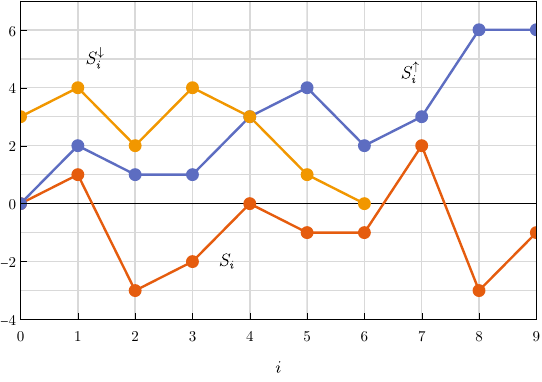}
    \caption{Illustration of the random walk $(S_i)_{i\geq 0}$ with increments distributed as $\nu_\mathbf{q}$, and its conditioned counterparts $(S_i^\downarrow)_{i\geq 0}$ and $(S_i^\uparrow)_{i\geq 0}$.}
    \label{fig:walks}
\end{figure}

Let us consider a random walk $(S_n)_{n\geq 0}$ with independent increments of law $\nu_\mathbf{q}$ (Fig.~\ref{fig:walks}). 
One may show \cite{Budd_Peeling_2016,Curien2019} that this random walk cannot have a positive drift\footnote{In case $\nu_\mathbf{q}$ has a finite first moment, a positive drift means $\mathbb{E}[S_{i+1}-S_i] = \sum_{k=-\infty}^\infty k \nu_\mathbf{q}(k) > 0$. More generally, a random walk has a positive drift if it has a non-zero probability of never visiting the negative integers.}.
By summing the probabilities \eqref{eq:perimeterlaw} over $k$, we deduce that $h^\downarrow$ is \emph{harmonic} on the positive integers with respect to $\nu_{\mathbf{q}}$, i.e.\ 
\begin{equation}\label{eq:nuharmonic}
    \sum_{k=-\infty}^\infty h^\downarrow(p+k)\nu_\mathbf{q}(k) = h^\downarrow(p) \qquad\text{for }p\geq 1.
\end{equation}
Since $h^\downarrow(p) \leq h^\downarrow(0) = 1$, the function $h^\downarrow$ acquires a simple probabilistic interpretation: when the random walk $(S_n)_{n\geq 0}$ is started at $p\geq 0$ and we consider the first time it visits the non-positive integers $\mathbb{Z}_{\leq 0}$, then $h^\downarrow(p)$ is the probability that it does so at $0$.
In the theory of random walks $h^\downarrow$ is said to be the \emph{pre-renewal function} of the random walk.
The transition probabilities \eqref{eq:perimeterlaw} then allow us to interpret the perimeter process $(P_i)_{i\geq 0}$ as having the law of the random walk $(S_n)_{n\geq 0}$ when conditioned on hitting $\mathbb{Z}_{\leq 0}$ at $0$, and killing it at that instance, which we denote by $(S_n^\downarrow)_{n\geq 0}$.
This conditioning with the help of a harmonic function is an example of a transformation of Markov chains known as Doob's $h$-transform \cite{Doob1957}.

In fact, the relation between admissible weight sequences and probability measures on $\mathbb{Z}$ is bijective \cite{Budd_Peeling_2016}, in the sense that any random walk on $\mathbb{Z}$ that has $h^\downarrow$ as its pre-renewal function has increments with law $\nu_\mathbf{q}$ for some admissible weight sequence $\mathbf{q}$.
This provides a different explanation for the universality observed in $\mathbf{q}$-Boltzmann maps\footnote{This could be made more explicit as follows \cite{Budd_Peeling_2017}. The Tutte equation \eqref{eq:tutte} is equivalent to $\nu_\mathbf{q}(-\ell-1) = \frac12 \sum_{p=-\infty}^\infty \nu_\mathbf{q}(p)\nu_{\mathbf{q}}(-p-k-1)$, which can be shown to imply that the law of the successive minima of $(S_i)_{i\geq 0}$ is universal, i.e.\ independent of $\mathbf{q}$. This in turn implies the universality of the probabilities $h^\downarrow(p)$ and thus of $W_\bullet^{(2\ell)}$.} and makes the measure $\nu_\mathbf{q}$ an economical alternative for $\mathbf{q}$ to specify a Boltzmann map model. 

\subsection{Peeling infinite Boltzmann planar maps}\label{sec:peelinginfinitemaps}

Note that \eqref{eq:nuharmonic} for $p=1$ is nothing but a disguised version of the admissibility criterion $g_\mathbf{q}(R)=1$ that we found in Section~\ref{sec:admissibilitycriticality}, because
\begin{equation}
    g_\mathbf{q}(R) -1 \stackrel{\eqref{eq:hdownandnu}}{=} 2R \left(h^\downarrow(1) - \sum_{k=-1}^\infty h^\downarrow(k+1)\nu_\mathbf{q}(k)\right).
\end{equation}
How about the criticality of $\mathbf{q}$? 
We claim that $\mathbf{q}$ is critical if the random walk $(S_n)_{n\geq 0}$ has no drift\footnote{More accurate terminology for ``no drift'' is that the random walk \emph{oscillates}, meaning that almost surely the range of the walk is unbounded above and below. In the case of finite first moments, this is equivalent to zero expectation value for the increments, $\sum_{k=-\infty}^\infty k \nu_\mathbf{q}(k) = 0$. However, the latter criterion is not very useful here, because in the next section we will see that for $\mathbf{q}$ subcritical $\nu_\mathbf{q}$ never has finite first moments.}, and thus is subcritical when it has negative drift.
To see this, we note that the criterion $g_\mathbf{q}'(R)=0$ for criticality from Section~\ref{sec:admissibilitycriticality} translates into
\begin{equation}\label{eq:crithup}
    0=g_\mathbf{q}'(R) = h^\uparrow(1) - \sum_{k=0}^\infty h^\uparrow(k+1)\nu_{\mathbf{q}}(k).
\end{equation}
where we have introduced the \emph{renewal function} (see \cite[Ch.~XII]{Feller1971} or \cite[App.~A]{Curien2019} for background on renewal theory of random walks)
\begin{equation}
    h^\uparrow(\ell) \coloneqq \sum_{p=0}^{\ell-1} h^\downarrow(p) = 2\ell\,h^\downarrow(\ell).
\end{equation}
Knowing that $h^\downarrow$ is harmonic on $\mathbb{Z}_{>0}$, the condition \eqref{eq:crithup} and therefore criticality of $\mathbf{q}$ is seen to be equivalent to $h^\uparrow$ being harmonic on $\mathbb{Z}_{>0}$ as well, meaning that it satisfies the same equation \eqref{eq:nuharmonic} with $h^\downarrow$ replaced by $h^\uparrow$.
But the renewal function of a random walk is harmonic if and only if it does not have negative drift \cite{Bertoin1994}. 
Since we already know that it cannot have positive drift, this verifies our claim.

Since $h^\uparrow$ is harmonic on the positive integers when $\mathbf{q}$ is critical, one may consider the $h^\uparrow$-transform of the random walk $(S_n)_{n\geq 0}$ resulting in the Markov chain $(S^\uparrow_i)_{i\geq 0}$ with transition probabilities
\begin{equation}
    \mathbb{P}( S_{i+1}^\uparrow = p+k | S_i^\uparrow = p) = \frac{h^\uparrow(p+k)}{h^\uparrow(p)} \nu_{\mathbf{q}}(k).
\end{equation}
Since $h^\uparrow(\ell) = 0$ for $\ell \leq 0$, the Markov chain $(S^\uparrow_n)_{n\geq 0}$ will never hit $\mathbb{Z}_{\leq 0}$, and can thus be interpreted as the random walk $(S_n)_{n\geq 0}$ \emph{conditioned to stay positive forever} \cite{Bertoin1994}.
Perhaps not surprisingly, this is exactly the law of the perimeter process of the infinite $\mathbf{q}$-Boltzmann planar map \cite{Budd_Peeling_2016}.
In other words, the targeted peeling explorations of a (critical) pointed $\mathbf{q}$-Boltzmann map and the exploration of its local limit are related by conditioning the perimeter process of the former to stay positive, by the transformation
\begin{equation}\label{eq:perimeterlawcomparison}
    \mathbb{P}_\infty^{(2\ell)}( P_{i+1} = m | P_i = p) = \frac{m}{p}\, \mathbb{P}_\bullet^{(2\ell)}( P_{i+1} = m | P_i = p).
\end{equation}

Let's try to understand why this is the case and at the same time construct the $\mathbf{q}$-IBPM as the limit of a Boltzmann planar map conditioned to have a large number $n$ of vertices.
From the definition in Section~\ref{sec:locallimits} we deduce that a necessary condition for local convergence is that for every map $\mathfrak{e}$ with a single hole we have the limit
\begin{equation}\label{eq:submapprobabilities}
    \lim_{n\to\infty}\mathbb{P}^{(2\ell)}_\bullet\big(\mathfrak{e}\subset \map \,\big| \,|\vertices(\map)| = n\big) = \mathbb{P}^{(2\ell)}_\infty(\mathfrak{e} \subset \map). 
\end{equation}
If $|\vertices(\mathfrak{e})| = k$ and the hole of $\mathfrak{e}$ has degree $2p$, then the probability on the left-hand side is given explicitly by
\begin{align}
    \mathbb{P}^{(2\ell)}_\bullet\big(\mathfrak{e}\subset \map \,\big| \,|\vertices(\map)| = n\big) &= w_{\mathbf{q}}(\mathfrak{e}) \frac{[t^{n-k}]W_\bullet^{(2p)}(t,\mathbf{q})}{[t^{n}]W_\bullet^{(2\ell)}(t,\mathbf{q})} \\
    &= w_{\mathbf{q}}(\mathfrak{e}) \frac{h^\downarrow(p)}{h^\downarrow(\ell)} 4^{p-\ell}\frac{[t^{n-k}]R(t,\mathbf{q})^{p}}{[t^{n}]R(t,\mathbf{q})^{\ell}}.
\end{align}
The asymptotics for $R(t,\mathbf{q})$ derived in Section \ref{sec:admissibilitycriticality} imply that the latter ratio approaches $\frac{p}{\ell} R^{p-\ell}$ as $n\to \infty$.
Hence, \eqref{eq:submapprobabilities} holds if and only if
\begin{equation}\label{eq:submapprobrelation}
    \mathbb{P}^{(2\ell)}_\infty(\mathfrak{e} \subset \map) = w_{\mathbf{q}}(\mathfrak{e}) (4R)^{p-\ell} \frac{h^\uparrow(p)}{h^\uparrow(\ell)} = \frac{p}{\ell} \mathbb{P}^{(2\ell)}_\bullet(\mathfrak{e} \subset \map).
\end{equation}
Specializing to the targeted peeling exploration, this in turn implies that the perimeter processes satisfy \eqref{eq:perimeterlawcomparison}.

In fact, these arguments give a convenient way of proving the local limit by an explicit construction of the $\mathbf{q}$-IBPM.
If we know the perimeter process $(P_i)_{i\geq 0}$ we can deduce from \eqref{eq:perimeterincrements} the sequence of events $\mathsf{C}_k$, $\mathsf{G}_{\star,k}$, $\mathsf{G}_{k,\star}$, flipping a coin to choose between the latter two, and construct the $\mathbf{q}$-IBPM by performing the peeling operations and filling in any holes with independent $\mathbf{q}$-Boltzmann maps.
The resulting infinite map is one-ended by construction and indeed satisfies \eqref{eq:submapprobrelation} (see \cite[Chapter~VII]{Curien2019} for details).

\subsection{Scaling limit of the perimeter process}\label{sec:perimscaling}

Having related the perimeter process of the pointed and infinite Boltzmann planar maps to conditioned random walks, we are in a good position to discuss scaling limits.
We first focus on the unconditioned random walk $(S_i)_{i\geq 0}$, because the conditioning of the walk via the $h$-transform should transfer to a similar conditioning of the continuous stochastic process in the limit.
Contrary to the random walks we encountered in Section~\ref{sec:crt} and \ref{sec:browniansphere}, the scaling limit will not be Brownian motion, because the probability measure $\nu_\mathbf{q}$ necessarily has infinite variance.

To see this, we need a better handle on the disk function $W^{(2\ell)}$, because it enters the negative half of the probability measure $\nu_\mathbf{q}$.
One way is to use \eqref{eq:Wginv} in combination with \eqref{eq:pointedevendisk}, which leads to the explicit expression
\begin{equation}\label{eq:diskintegral}
    W^{(2\ell)} = \int_0^1 \rmd t\,W_\bullet^{(2\ell)}(t,\mathbf{q}) = \binom{2\ell}{\ell} \int_0^1 \rmd t R(t,\mathbf{q})^\ell = \binom{2\ell}{\ell} \int_0^R \rmd r\,g_\mathbf{q}'(r) r^\ell.
\end{equation}
When $\mathbf{q}$ is subcritical, $g_\mathbf{q}'(R)>0$, we read off that as $\ell\to\infty$ we have the asymptotics
\begin{equation}
    W^{(2\ell)} \sim  \binom{2\ell}{\ell} g_\mathbf{q}'(R) \frac{R^{\ell+1}}{\ell+1} \sim \frac{\mathsf{p}_\mathbf{q}}{2}(4R)^{\ell+1} \ell^{-3/2},\quad \mathsf{p}_\mathbf{q} = \frac{g_\mathbf{q}'(R)}{2\sqrt{\pi}},
\end{equation}
while in the generic critical ($a=\tfrac52$) and non-generic critical case ($\tfrac32<a<\tfrac52$) with expansion $g_\mathbf{q}(r) = 1 - C (R-r)^{a-1/2} + o((R-r)^{a-\tfrac12})$ we deduce that
\begin{equation}
    W^{(2\ell)} \sim  \binom{2\ell}{\ell} C R^{\ell+a-\tfrac12} \ell^{\tfrac12-a} \Gamma(a+\tfrac12)\sim\frac{\mathsf{p}_\mathbf{q}}{2} (4R)^{\ell+1} \ell^{-a}, \quad \mathsf{p}_\mathbf{q}= \frac{C\Gamma(a+\tfrac12)R^{a-\tfrac32}}{4\sqrt{\pi}}.
\end{equation}
Hence, if we set $a=3/2$ in the subcritical case, then we can summarize the negative tail behaviour as $\nu(-k) \sim \mathsf{p}_\mathbf{q} k^{-a}$.
The positive tail can be shown \cite{Borot_recursive_2012,Budd_Peeling_2016,Budd_Geometry_2017,Curien2019} to be negligible in the subcritical and generic critical case, while in the non-generic critical case it is asymptotically smaller by a factor of $\cos a\pi$, as summarized in Table~\ref{tab:walkscaling} below.

\begin{table}[!h]
    \caption{Scaling properties of the random walk $(S_i)_{i\geq0}$ depending on type of weight sequence.}
    \label{tab:walkscaling} 
    \renewcommand\arraystretch{1.5}
    \renewcommand\tabcolsep{2.5mm}
    \begin{tabular}{cccc}
    \hline\noalign{\smallskip} 
    Type & Subcritical  & Generic critical & Non-generic critical ($\tfrac32<a<\tfrac52$)  \\
    \noalign{\smallskip}\svhline\noalign{\smallskip}
    definition &$g'(R) > 0$ & $\begin{array}{c} g'(R) = 0,\\\,|g''(R)| < \infty\end{array}$ & $1-g(r) \stackrel{r\to R}{\sim} C(R-r)^{a-\tfrac12}$\\
    drift of $(S_n)_{n\geq 0}$& negative & no drift & no drift \\
    $\nu_\mathbf{q}(-k)$ & $\stackrel{k\to\infty}{\sim}  \mathsf{p}_\mathbf{q} k^{-\tfrac32}$ & $\stackrel{k\to\infty}{\sim} \mathsf{p}_\mathbf{q} k^{-\tfrac52}$ & $\stackrel{k\to\infty}{\sim}  \mathsf{p}_\mathbf{q} k^{-a}$ \\
    ${\displaystyle\lim_{k\to\infty}\frac{\sum_{\ell=k}^\infty \nu_\mathbf{q}(\ell)}{\sum_{\ell=k}^\infty \nu_\mathbf{q}(-\ell)}}$ & $0$ & $0$ & $\cos a\pi$ \\
    \noalign{\smallskip}\hline
    \end{tabular}
\end{table}

\begin{figure}[t]
    \centering
    \sidecaption[t]
    \includegraphics[width=.6\linewidth]{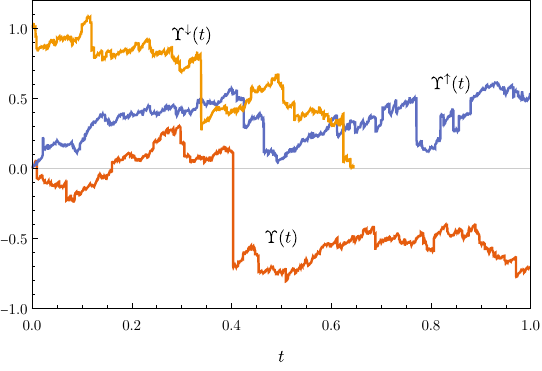}
    \caption{Simulations of the self-similar processes $\Upsilon(t)$, $\Upsilon^\downarrow(t)$ and $\Upsilon^\uparrow(t)$ started at $\Upsilon(0) = \Upsilon^\uparrow(0)=0$ and $\Upsilon^\downarrow(0)=1$ in the case $a=2.4$.
    They correspond to a L\'evy stable process with L\'evy measure \eqref{eq:levymeasure} and its counterparts that are conditioned to die continuously at $0$, respectively, stay positive forever.}
    \label{fig:upsplot}
\end{figure}

Based on the tails of the distribution, we should be looking at convergence of $(S_i)_{i\geq0}$ to a \emph{L\'evy stable process} $\left(\Upsilon(t)\right)_{t\geq 0}$ with no drift or Brownian component but with \emph{L\'evy measure}
\begin{equation}\label{eq:levymeasure}
    \frac{\rmd x}{|x|^a} \ind_{\{x<0\}} + \cos a\pi \frac{\rmd x}{|x|^a} \ind_{\{x>0\}},
\end{equation} 
which informally expresses the exponential rate at which the process performs jumps of size $x \in \mathbb{R}$. 
See Fig.~\ref{fig:upsplot} for a simulations.
Like Brownian motion, its increments are independent and stationary\footnote{Meaning that for any $s > 0$, the increments $\Upsilon(is)-\Upsilon((i-1)s)$, $i=1,2,\ldots$, are independent and identically distributed.}, but the sample paths are not continuous and it satisfies a scaling relation with an exponent $a-1\neq 2$,
\begin{equation}\label{eq:scalingstable}
    \left(\lambda^{-1}\Upsilon(\lambda^{a-1} t)\right)_{t\geq 0} \stackrel{\mathrm{(d)}}{=} \big(\Upsilon(t)\big)_{t\geq 0}, \qquad \lambda>0.
\end{equation}

According to the generalized central limit theorem \cite{Gnedenko_Limit_1954}, the sum $S_n$ of $n$ random integers with distribution $\nu_\mathbf{q}$ converges to the stable random variable\footnote{Once the tails of $\nu_\mathbf{q}$ are known, the only non-trivial check to perform is that no centering of the sequence $n^{-1/(a-1)}S_n$ by an $n$-dependent shift is required for convergence. For $a\neq 2$ this is straightforward, while the case $a=2$ requires some care \cite{Budd2019}.}
\begin{equation}
    \frac{S_n}{n^{1/(a-1)}} \xrightarrow[n\to\infty]{\mathrm{(d)}} \Upsilon_a(\mathsf{p}_\mathbf{q}).
\end{equation}
With an appropriate choice of topology on non-continuous sample paths (named after Skorokhod) this convergence extends to the full process \cite{Jacod_Limit_2003},
\begin{equation}\label{eq:walkconvergence}
    \left( \frac{S_{\lfloor\lambda t/\mathsf{p}_\mathbf{q}\rfloor}}{\lambda^{1/(a-1)}} \right)_{t\geq0} \xrightarrow[\lambda\to\infty]{\mathrm{(d)}} \left( \Upsilon(t) \right)_{t\geq0}.
\end{equation}
Due to an invariance principle of \cite{Caravenna_Invariance_2008}, one obtains a similar convergence of the random walk $(S_i^\uparrow)_{i\geq0}$ conditioned to stay positive, and therefore of the perimeter process $(P_i)_{i\geq 0}$ of the $\mathbf{q}$-IBMP which has the same law, 
\begin{equation}
    \left( \frac{P_{\lfloor\lambda t/\mathsf{p}_\mathbf{q}\rfloor}}{\lambda^{1/(a-1)}} \right)_{t\geq0} \xrightarrow[\lambda\to\infty]{\mathrm{(d)}} \left( \Upsilon^\uparrow(t) \right)_{t\geq0}.
\end{equation}
Here $\Upsilon^\uparrow(t)$ is the stable process started at $0$ and conditioned to stay positive for all $t>0$.
Note that in both convergences the starting point, $S_0$ respectively $P_0$, is kept fixed.

In the case of the conditioned random walk $(S^\downarrow_n)_{n\geq 0}$ started at $S_0 = \ell$, one should instead consider the limit $\ell \to \infty$ and rescale time accordingly.
So for the pointed $\mathbf{q}$-Boltzmann planar map with perimeter $2\ell$ we have, with the help of another invariance principle of \cite{Caravenna_Invariance_2008}, the convergence
\begin{equation}\label{eq:perimeterscalingpointed}
    \left( \frac{P_{\lfloor\ell^{a-1}t/\mathsf{p}_\mathbf{q}\rfloor}}{\ell} \right)_{t\geq0} \xrightarrow[\ell\to\infty]{\mathrm{(d)}} \left( \Upsilon^\downarrow(t) \right)_{t\geq0},
\end{equation}
where $\Upsilon^\downarrow(t)$ is the stable process $\Upsilon(t)$ conditioned \emph{to die continuously at $0$}, meaning that it is killed at time $\tau = \inf \{ t: \Upsilon(t)\leq 0\}$ and conditioned to do so continuously, $\lim_{t\nearrow \tau} \Upsilon(t) = 0$.
The processes $\Upsilon^\uparrow(t)$ and $\Upsilon^\downarrow(t)$ are examples of \emph{(positive) self-similar Markov processes} (see e.g. \cite{Kyprianou_Fluctuations_2014}), meaning that, even though the increments are no longer independent or stationary, they satisfy the same scaling relation \eqref{eq:scalingstable}, with the understanding that if the process $\Upsilon^\downarrow$ on the left-hand side starts at $\Upsilon^\downarrow(0)=x$ then the one on the right-hand side starts at $\Upsilon^\downarrow(0)=x/\lambda$.

A consequence of \eqref{eq:perimeterscalingpointed} is that in the large-$\ell$ limit it takes the targeted peeling exploration on the order of $\ell^{a-1}$ steps to find the target.
This is much smaller than the total number of faces and vertices in the map, which is of order $\ell^{a-1/2}$.
Indeed, it is straightforward to compute the expected number of vertices in a (unpointed) $\mathbf{q}$-Boltzmann planar map with root face degree $2\ell$ since it is given by the ratio of the pointed and unpointed disk function,
\begin{equation}
    \mathbb{E}_\mathbf{q}[|\vertices(\map)|] = \frac{W_\bullet^{(2\ell)}}{W^{(2\ell)}} = \frac{h^\downarrow(\ell)}{4R\nu_\mathbf{q}(-\ell-1)} \sim \mathsf{b}_\mathbf{q} \ell^{a-\tfrac12}, \qquad \mathsf{b}_\mathbf{q} = \frac{1}{2R\mathsf{p}_\mathbf{q} \sqrt{\pi}}.
\end{equation}
More precisely, based on singularity analysis of $W^{(2\ell)}$ in \eqref{eq:diskintegral}, one may verify that there exists a real random variable $\xi$ of mean $1$ such that \cite{Budd_Geometry_2017}
\begin{equation}
    \frac{|\vertices(\map)|}{\mathsf{b}_\mathbf{q} \ell^{a-\tfrac12}} \xrightarrow[\ell\to\infty]{\mathrm{(d)}} \xi,\qquad \mathbb{E}[\xi e^{-\lambda \xi}] = \exp\left( - \left[\Gamma(a+\tfrac12)\lambda\right]^{\frac{1}{a-1/2}}\right).
\end{equation}
Letting $|\vertices(\bar{\mathfrak{e}}_i)|$ be the number of explored vertices after $i$ steps in the targeted peeling exploration, then $|\vertices(\bar{\mathfrak{e}}_{i+1})|-|\vertices(\bar{\mathfrak{e}}_i)|$ is non-zero only in the gluing event $\mathsf{G}_{\star,k}$ or $\mathsf{G}_{k,\star}$.
In this case it is the number of vertices of an independent $\mathbf{q}$-Boltzmann map with root face degree $2k$. 
It should therefore not come as a surprise that $|\vertices(\bar{\mathfrak{e}}_i)|$ possesses a scaling limit that jumps whenever $S^\uparrow$ makes a negative jump \cite{Curien2017,Budd_Geometry_2017},
\begin{equation}\label{eq:perimvolumescaling}
    \left( \frac{P_{\lfloor\lambda t/\mathsf{p}_\mathbf{q}\rfloor}}{\lambda^{1/(a-1)}}, \frac{|\vertices(\bar{\mathfrak{e}}_{\lfloor\lambda t/\mathsf{p}_\mathbf{q}\rfloor})|}{\mathsf{b}_\mathbf{q}\lambda^{\frac{a-1/2}{a-1}}}\right) \xrightarrow[\ell\to\infty]{\mathrm{(d)}} \left(\Upsilon^\uparrow(t),\mathcal{V}(t)\right)_{t\geq 0}.
\end{equation}
Here $\mathcal{V}(t)$ is the increasing stochastic process defined as follows.
Since the jumps of $\Upsilon^\uparrow$ are countable, we can list the times of negative jumps as $t_1,t_2,\ldots$ and denote their magnitudes by $\Delta_i = \lim_{\epsilon\nearrow 0} \Upsilon^\uparrow(t_i-\epsilon)-\Upsilon^\uparrow(t_i+\epsilon)$.
If $\xi_1,\xi_2,\ldots$ are independent random variables with the distribution of $\xi$ then $\mathcal{V}(t)$ is given by
\begin{equation}
    \mathcal{V}(t) = \sum_{i=1}^\infty \xi_i \,\Delta_i^{a-\tfrac{1}{2}} \ind_{\{t_i <t\}}.
\end{equation}
Observe that $\left(\Upsilon^\uparrow(t),\mathcal{V}(t)\right)_{t\geq 0}$ is exactly self-similar again, in the sense that
\begin{equation}
    \left(\lambda^{-1}\Upsilon^\uparrow(\lambda^{a-1}t),\lambda^{a-1/2}\mathcal{V}(\lambda^{a-1}t)\right)_{t\geq 0} \stackrel{\mathrm{(d)}}{=} \left(\Upsilon^\uparrow(t),\mathcal{V}(t)\right)_{t\geq 0}.
\end{equation}

\subsection{Geometry}\label{sec:peelinggeometry}

So far we have not specified a peeling algorithm $\mathcal{A}$, since the law of the perimeter process and the collection of Boltzmann maps that fill in the holes were independent of this choice.
We may thus design the algorithm to suit whatever application we have in mind.
To connect with the study or the distances discussed in previous sections, we will focus on explorations in infinite $\mathbf{q}$-Boltzmann planar maps that correspond to balls of growing radius around the root.
Contrary to the tree bijections that rely on the graph distance on the map, the peeling exploration is most naturally formulated in terms of distances on its dual.

\begin{figure}[t]
    \centering
    \includegraphics[width=\linewidth]{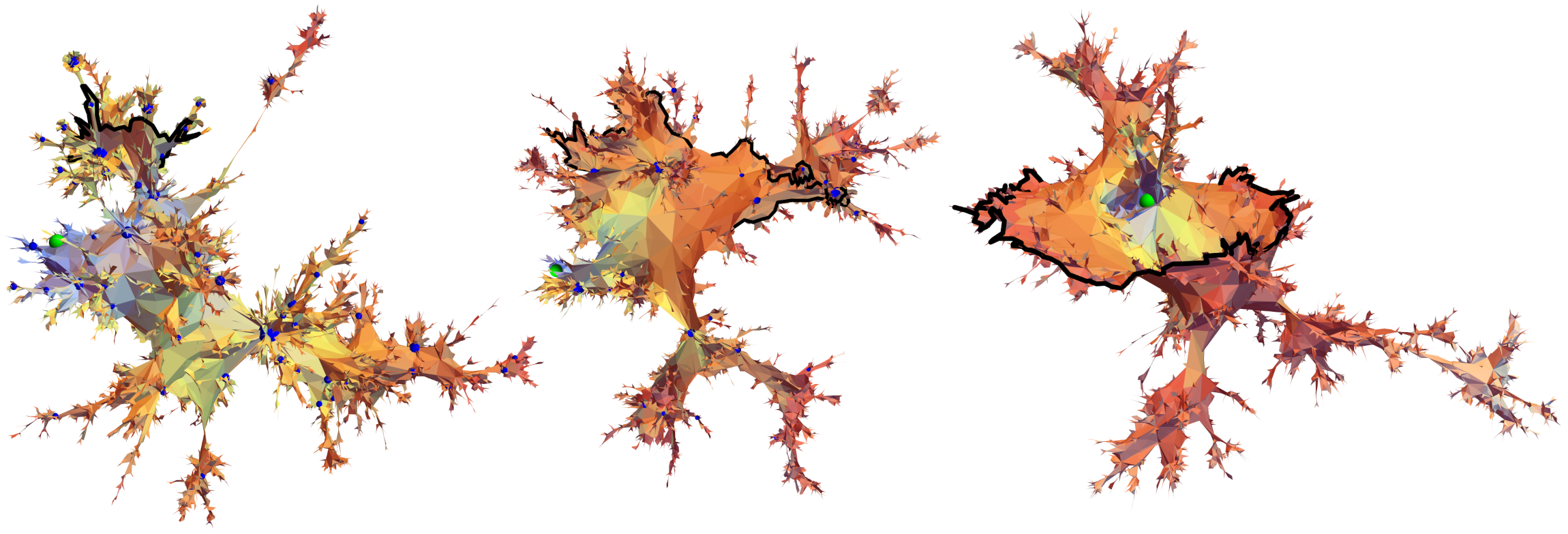}
    \caption{Simulations of the hull $\overline{\mathsf{Ball}}_r^\dagger(\map)$ of radius $r \approx 20$ in infinite $\mathbf{q}$-Boltzmann maps of type $a=2.1,\, 2.25,\,2.5$ (left to right). The boundary of the hull (of length $L_r$) is drawn in black, and the root as a green sphere. Blue spheres indicate faces of $\map$ of high degree.}
    \label{fig:hulls}
\end{figure}

\begin{figure}[t]
    \centering
    \includegraphics[width=.8\linewidth]{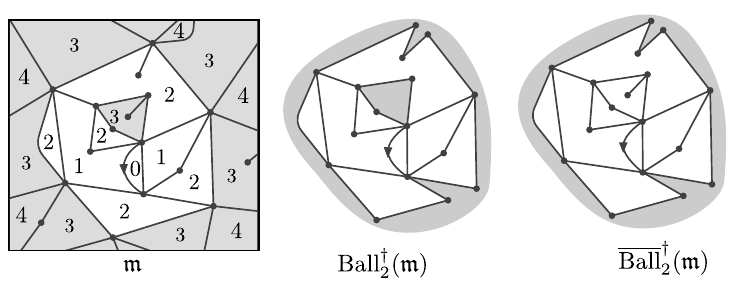}
    \caption{A portion of an infinite map $\map$ is shown with its faces labeled by dual graph distance to the root face (the face to the left of the root edge). The submaps corresponding to the dual ball of radius $2$ and its hull are shown on the right.}
    \label{fig:dualball}
\end{figure}

To be precise, the \emph{dual graph distance} $d^\dagger(f,f')$ between faces $f$ and $f'$ of a map $\map$ is given by fewest number of hops between adjacent faces necessary to get from $f$ to $f'$. 
We then let the \emph{dual ball of radius $r$} in $\map$ be the submap $\mathsf{Ball}_r^\dagger(\map) \subset \map$ given by keeping all faces of $\map$ that are at dual graph distance at most $r$ from the root face, but cutting open every edge shared by two faces at distance exactly $r$.
In addition, we consider its \emph{hull} $\overline{\mathsf{Ball}}_r^\dagger(\map) \subset \map$ to be the ball $\mathsf{Ball}_r^\dagger(\map) \subset \map$ with all finite holes filled in.
See Fig.~\ref{fig:hulls} for some simulations of hulls $\overline{\mathsf{Ball}}_r^\dagger(\map)$ and Fig.~\ref{fig:dualball} for an illustration.

The reason for these particular definitions is that we can choose the peeling algorithm $\mathcal{A}$ in such a way that all hulls $\overline{\mathsf{Ball}}_r^\dagger(\map)$, $r \geq 0$, occur in the targeted peeling exploration, 
\begin{equation}\label{eq:ballexploration}
    \bar{\mathfrak{e}}_{\theta_r} = \overline{\mathsf{Ball}}_r^\dagger(\map) 
\end{equation}
for some increasing sequence of indices $0=\theta_0 < \theta_1 < \theta_2 < \cdots$.

This is achieved by taking $\mathcal{A}(\bar{\mathfrak{e}}_i)$ to return an active edge of minimal height \cite{Budd_Geometry_2017}, where we define the \emph{height} of an active edge $e$ to be the dual graph distance between the root face and the face of $\bar{\mathfrak{e}}_i$ that is adjacent to $e$.
We denote this minimal height by $H_i$.
One may verify inductively that at any time in such an exploration all active edges in $\bar{\mathfrak{e}}_i$ are adjacent to a face at distance $H_i$ or $H_i+1$ and that \eqref{eq:ballexploration} holds for
\begin{equation}
    \theta_r = \inf \{ i : H_i \geq r\}.
\end{equation}
To specify $\mathcal{A}$ more precisely, one could choose to explore layer by layer in a clockwise fashion, meaning that one takes $\mathcal{A}(\bar{\mathfrak{e}}_i)$ to be the active edge at height $H_i$ that sits just to the right of one at height $H_i+1$ (if it exists).
This ensures that the active edges at height $H_i$ always make up a connected portion of the boundary. 

We can now apply the scaling limit results of Section~\ref{sec:perimscaling} to analyze the growth of the boundary length and volume of the geodesic ball around the root of increasing radius (as illustated in Fig.~\ref{fig:peelingillustration}). 
To be precise, we set $L_r$ to be half of the degree of the hole of the dual ball hull $\overline{\mathsf{Ball}}_r^\dagger(\map)$ and $V_r$ its number of vertices.
By our previous considerations, $L_r$ and $V_r$ are related to the perimeter process $(P_i)_{i\geq 0}$ via 
\begin{equation}\label{eq:LVtoperim}
    L_r = P_{\theta_r}, \qquad V_r = |\vertices(\bar{\mathfrak{e}}_{\theta_r})|,
\end{equation}
so we need to get a handle on $\theta_r$, in particular on the number of peeling steps $\Delta\theta_r = \theta_{r+1} - \theta_r$ required to explore all faces at dual graph distance $r+1$.

At time $\theta_r$ there are precisely $2p = 2P_{\theta_r}$ active edges at height $r$, so our first guess would be that $\Delta\theta_r  \approx 2p$.
This would be the exact answer if each of the peeling steps would uncover a new face (event $\mathsf{C}_k$), necessarily at dual graph distance $r+1$.
However, this is accelerated by gluing events $\mathsf{G}_{\star,k}$ that will typically swallow $k$ additional active edges at height $r$.
The probability of $\mathsf{G}_{\star,k}$ in the large-$p$ limit is $\tfrac12 \nu_\mathbf{q}(-k-1) \sim \mathsf{p}_\mathbf{q} k^{-a}$, so whether this acceleration changes the scaling depends on  the type $a$ \cite{Budd_Geometry_2017,Budd2019}. 
If $a > 2$ then each step swallows on average a finite number of edges, so $\Delta\theta_r$ is still proportional to the perimeter $2p$; if $a\leq 2$ then edges are swallowed fast enough that the scaling changes and $\Delta\theta_r$ becomes of the order $p / \log p$ for $a=2$ and $p^{a-1}$ for $a<2$.
In terms of the height $H_n$ after $n$ steps, this translates into the approximate growth
\begin{equation}\label{eq:heightgrowth}
    H_n \approx \begin{cases}\sum_{i=1}^n \frac{1}{P_i} \quad\approx n^{\frac{a-2}{a-1}}& \text{for }a\in (2,\tfrac52], \\ \sum_{i=1}^n \frac{\log{P_i}}{P_i}\approx \log^2 n & \text{for }a=2, \\ \sum_{i=1}^n P_i^{1-a} \mkern1mu\approx \log n & \text{for }a \in (\tfrac32,2),\end{cases}
\end{equation}
where we used that $P_n \approx n^{\frac{1}{a-1}}$ as $n\to\infty$.

For $a \leq 2$ we conclude that $\theta_r$ grows faster than any power law in $r$, and the same therefore holds for $L_r$ and $V_r$.
In fact, one may show \cite{Budd_Geometry_2017,Budd2019} rigorously that as $r\to\infty$ we have the convergence in probability
\begin{align}
    \log L_r &\sim \pi \sqrt{2r}, \qquad \log V_r \sim \pi \frac{3\pi}{\sqrt{2}}\sqrt{r}\qquad\mkern15mu \text{for }a=2\\
    \log L_r & \sim c r, \qquad \mkern25mu \log V_r \sim (a-\tfrac12)c r \qquad\text{for }a\in(\tfrac32,2).
\end{align}
This implies that the random metric spaces obtained from the dual graph distance on the vertex set of (pointed or infinite) $\mathbf{q}$-Boltzmann maps that are non-generic critical of type $a\leq 2$ do \emph{not} possess scaling limits.
Informally, one could say that they correspond to a pathological situation of infinite Hausdorff dimension.

In the generic critical case $a=\tfrac52$ or the non-generic critical case $a\in(2,\tfrac52)$ we see that $\theta_r \approx n^{\frac{a-1}{a-2}}$ and therefore $L_r = P_{\theta_r} \approx r^{\frac{1}{a-2}}$. 
The estimate \eqref{eq:heightgrowth} can be justified and turned into a scaling limit jointly with that of the perimeter process \cite{Budd_Geometry_2017}: there exists a $c>0$ such that
\begin{equation}
    \left(c \lambda^{\frac{a-1}{a-2}} H_{\lfloor \lambda t \rfloor}\right)_{t\geq 0} \xrightarrow[\lambda\to\infty]{\mathrm{(d)}} \left(\int_0^t \frac{\rmd u}{\Upsilon^\uparrow(u)}\right)_{t\geq0},
\end{equation}
and the right-hand side can be shown to be finite almost surely.
Combining with \eqref{eq:perimvolumescaling} and \eqref{eq:LVtoperim}, this implies that $L_r$ and $V_r$ scale towards a reparametrized version of the Markov process $(\Upsilon^\uparrow(t),\mathcal{V}(t))$,
\begin{align}
    \left( \frac{L_{\lfloor \lambda x/(c\mathsf{p}_\mathbf{q})\rfloor}}{\lambda^{\frac{1}{a-2}}},\frac{V_{\lfloor \lambda x/(c\mathsf{p}_\mathbf{q})\rfloor}}{\mathsf{b}_\mathbf{q}\lambda^{\frac{a-1/2}{a-2}} } \right)_{x\geq 0} &\xrightarrow[\lambda\to\infty]{\mathrm{(d)}} \left(\Upsilon^\uparrow(\vartheta(x)), \mathcal{V}(\vartheta(x))\right)_{x\geq0},\\
    \vartheta(x)\coloneqq &\inf\left\{ t : \int_0^t \frac{\rmd u}{\Upsilon^\uparrow(u)} \geq x\right\}.
\end{align}
Although the limit may not feel like a very tangible stochastic process, it is universal in the sense that its distribution only depends on the type $a$.
In the generic case $a=\tfrac52$, it corresponds to the boundary length and area of hulls of geodesic radius $x$ in the Brownian plane \cite{Curien2016}, and explicit formulae for the distribution of $\Upsilon^\uparrow(\vartheta(x))$ and $\mathcal{V}(\vartheta_x)$ can be derived using various approaches \cite{Krikun2005,Menard2016,Curien2016,Curien2017}.

In general, it follows from the self-similarity relation
\begin{equation}
    \left(\lambda^{- \frac{1}{a-2}}\Upsilon^\uparrow(\vartheta(\lambda x),\lambda^{- \frac{a-1/2}{a-2}}\mathcal{V}(\vartheta(\lambda x))\right)_{x\geq 0} \stackrel{\mathrm{(d)}}{=} \left(\Upsilon^\uparrow(\vartheta(x)),\mathcal{V}(\vartheta(x))\right)_{x\geq 0},
\end{equation}
that a potential scaling limit in the Gromov--Hausdorff sense must have a Hausdorff dimension
\begin{equation}
    d_{\mathrm{H}} = \frac{a-\tfrac12}{a-2}.
\end{equation}
In the generic critical case $a=5/2$ this reproduces the Hausdorff dimension $d_{\mathrm{H}} = 4$ from the construction of the Brownian sphere in Section~\ref{sec:browniansphere}. 
For $a\in(2,\tfrac52)$, it is conjectured that the scaling limits correspond to a new family of universality classes of random metrics on the $2$-sphere, tentatively referred to as the \emph{stable spheres}, but a full construction of these random metrics is still out of reach \cite{Budd_Geometry_2017,Bertoin_Martingales_2017}. 

These metric spaces are quite different from the ones obtained by examining the normal graph distance in non-generic critical maps of type $a \in (\tfrac32,\tfrac52)$.
Gromov--Hausdorff limits (at least along subsequences) of the latter have been obtained in \cite{LeGall_Scaling_2011} and are referred to as \emph{stable maps}.
They have Hausdorff dimension $d_{\mathrm{H}} = 2a-1$ and do not have the topology of the 2-sphere, but contain macroscopic holes (arising from faces of macroscopic degree in the limit).

\begin{acknowledgement}
This work is supported by the START-UP 2018 programme with project number 740.018.017 and the VIDI programme with project number VI.Vidi.193.048, which are
financed by the Dutch Research Council (NWO).
\end{acknowledgement}

\providecommand{\href}[2]{#2}\begingroup\raggedright\endgroup


\end{document}